\documentclass[12pt]{article}
\pdfoutput=1
\usepackage[utf8]{inputenc}
\usepackage{epsf}
\usepackage{colortbl}
\usepackage{amsmath}
\usepackage{amsfonts}
\usepackage{amssymb}
\usepackage{graphicx}
\usepackage{color}
\usepackage{psfrag}
\usepackage{cite}
\usepackage{hyperref}
\usepackage{tikz}
\usepackage[normalem]{ulem}

\usepackage{ifpdf}

\newcommand{\bmat}{\left(\begin{array}}
\newcommand{\emat}{\end{array}\right)}

\def\yzero{\smash{\hbox{$y\kern-4pt\raise1pt\hbox{${}^\circ$}$}}}

\def\beq{\begin{equation}}
\def\eeq{\end{equation}}
\def\beqa{\begin{eqnarray}}
\def\eeqa{\end{eqnarray}}

\def\-{\hphantom{-}}

\def\s2{\frac{1}{\sqrt2}}

\def\beq{\begin{equation}}
\def\eeq{\end{equation}}
\def\beqa{\begin{eqnarray}}
\def\eeqa{\end{eqnarray}}

\def\IF{\relax{\rm I\kern-.18em F}}
\def\II{\relax{\rm I\kern-.18em I}}

\def\Dsl{\,\raise.15ex\hbox{/}\mkern-13.5mu D} %this one can be subscripted
\newcommand{\eq}[1]{(\ref{#1})}
\newcommand{\ket}[1]{\vert #1 \rangle}

\catcode`\@=11   
\newdimen\@rotdimen
\newbox\@rotbox  

\def\@vspec#1{\special{ps:#1}}%  passes #1 verbatim to the output
\def\@rotstart#1{\@vspec{gsave currentpoint currentpoint translate
   #1 neg exch neg exch translate}}% #1 can be any origin-fixing transformation
\def\@rotfinish{\@vspec{currentpoint grestore moveto}}% gets back in synch 
\def\@rotr#1{\@rotdimen=\ht#1\advance\@rotdimen by\dp#1%
   \hbox to\@rotdimen{\hskip\ht#1\vbox to\wd#1{\@rotstart{90 rotate}%
   \box#1\vss}\hss}\@rotfinish}
\def\@rotl#1{\@rotdimen=\ht#1\advance\@rotdimen by\dp#1%
   \hbox to\@rotdimen{\vbox to\wd#1{\vskip\wd#1\@rotstart{270 rotate}%
   \box#1\vss}\hss}\@rotfinish}%
\def\@rotu#1{\@rotdimen=\ht#1\advance\@rotdimen by\dp#1%
   \hbox to\wd#1{\hskip\wd#1\vbox to\@rotdimen{\vskip\@rotdimen
   \@rotstart{-1 dup scale}\box#1\vss}\hss}\@rotfinish}%
\def\@rotf#1{\hbox to\wd#1{\hskip\wd#1\@rotstart{-1 1 scale}%
   \box#1\hss}\@rotfinish}%
\def\rotate{\@ifnextchar[{\@rotate}{\@rotate[l]}}
\def\@rotate[#1]#2{\setbox\@rotbox=\hbox{#2}\@nameuse{@rot#1}\@rotbox}

\catcode`\@=12
\topmargin
-1.5cm
\textwidth
15.5cm
\textheight
23.5cm
\oddsidemargin
0.7cm
\evensidemargin
1.2cm

\begin{document}

%----------------------------------------------------------------------%
%  numbering equations with section number
%----------------------------------------------------------------------%
\makeatletter
\@addtoreset{equation}{section}
\makeatother
\renewcommand{\theequation}{\thesection.\arabic{equation}}
%----------------------------------------------------------------------%
%  title page
%----------------------------------------------------------------------%
\hypersetup{pageanchor=false}
\pagestyle{empty}
%\vspace*{1.0in}
\rightline{ IFT-UAM/CSIC-16-060}
 \rightline{ FTUAM-16-25}
\rightline{MAD-TH-16-03}

\vspace{1.2cm}
\begin{center}
\LARGE{\bf The Weak Gravity Conjecture in three dimensions \\[12mm] }
\large{Miguel Montero$^{1,2}$,  Gary Shiu$^{3,4}$, Pablo Soler$^{3,4}$\\[4mm]}
\footnotesize{${}^{1}$ Departamento de F\'{\i}sica Te\'orica, Facultad de Ciencias\\[-0.3em] 
Universidad Aut\'onoma de Madrid, 28049 Madrid, Spain\\
${}^2$ Instituto de F\'{\i}sica Te\'orica IFT-UAM/CSIC,\\[-0.3em] 
C/ Nicol\'as Cabrera 13-15, 
Campus de Cantoblanco, 28049 Madrid, Spain} \\ 
${}^3$ {Department of Physics, University of Wisconsin-Madison, \\ Madison, WI 53706, USA} \\
${}^4$ { Department of Physics \& Institute for Advanced Study,\\ Hong Kong University of Science and Technology, Clear Water Bay, Hong Kong}\\

\vspace*{5mm}

\small{\bf Abstract} \\%[5mm]
\end{center}
\begin{center}
\begin{minipage}[w]{12.3cm}
We study weakly coupled $U(1)$ theories in $AdS_3$, their associated charged BTZ solutions, and their charged spectra. We find that modular invariance of the holographic dual two-dimensional CFT and compactness of the gauge group together imply the existence of charged operators with conformal dimension significantly below the black hole threshold. We regard this as a form of the Weak Gravity Conjecture (WGC) in three dimensions. We also explore the constraints posed by modular invariance on a particular discrete $\mathbb{Z}_N$ symmetry which arises in our discussion. In this case, modular invariance does not guarantee the existence of light $\mathbb{Z}_N$-charged states. We also highlight the differences between our discussion and the usual heuristic arguments for the WGC based on black hole remnants.
\end{minipage}
\end{center}
\newpage
\hypersetup{pageanchor=true}
%----------------------------------------------------------------------%
%  Resetting of counters
%----------------------------------------------------------------------%
\setcounter{page}{1}
\pagestyle{plain}
\renewcommand{\thefootnote}{\arabic{footnote}}
\setcounter{footnote}{0}
%----------------------------------------------------------------------%
%  Paper begins
%----------------------------------------------------------------------%

\tableofcontents 

\vspace*{1cm}

\section{Introduction}
The Weak Gravity Conjecture (WGC)~\cite{ArkaniHamed:2006dz} predicts the existence of light charged objects in any weakly coupled gauge theory also coupled to gravity, and has been a very useful tool in understanding the constraints that quantum gravity imposes on effective field theories. In particular, it has been recently applied to large-field inflationary models and relaxions \cite{Rudelius:2014wla,delaFuente:2014aca,Rudelius:2015xta,Montero:2015ofa,Brown:2015iha,Bachlechner:2015qja,Hebecker:2015rya,Brown:2015lia,Junghans:2015hba,Heidenreich:2015wga,Palti:2015xra,Heidenreich:2015nta,Kooner:2015rza,Kappl:2015esy,Choi:2015aem, Ibanez:2015fcv,Hebecker:2015zss,Fonseca:2016eoo,Parameswaran:2016qqq,Baume:2016psm,Garcia-Valdecasas:2016voz}. 

The power of the WGC lies in its generality. Relying solely in low energy data, it can be used to constrain effective field theories without the need to refer to particular UV embeddings, or scan through the string landscape. Thus, if correct, it is a most useful tool to explore the Swampland~\cite{Vafa:2005ui}.

Although, to our knowledge, the WGC has been verified in every string theory example considered so far (in particular in simple heterotic setups~\cite{ArkaniHamed:2006dz}), a general proof of the conjecture is still missing. Already in the original paper, a heuristic argument for the conjecture was given, essentially claiming that the conjecture is necessary to prevent the local gauge symmetry from becoming global, which has its own problems with quantum gravity (see e.g. \cite{Susskind:1995da,Banks:2010zn}). This was made more precise in \cite{Banks:2006mm}, but the arguments have loopholes and we do not see a clear way to turn them into a formal proof. A promising avenue was opened by \cite{Nakayama:2015hga} (see also \cite{Harlow:2015lma} and \cite{Horowitz:2016ezu}), which studied the WGC in the context of the AdS/CFT correspondence,  where it can be translated into a statement about the CFT which can then be verified or disproved. Reference \cite{Nakayama:2015hga} found certain examples of CFT's which seem not to satisfy the WGC. However, it is not clear that these CFT's have weakly coupled gravitational duals, so that the WGC might actually still be correct; alternatively, maybe the WGC only holds for a restricted class of CFT's.

In this paper, we study the WGC in three-dimensional AdS space. On one hand, the drastic differences in the behavior of both gravity and gauge fields between three and more dimensions may mean that this problem is unrelated to higher dimensional cases. On the other hand, the greatly enhanced conformal symmetry of the two-dimensional holographic dual greatly simplifies the problem, and in three dimensions we do have the essential ingredients of the WGC (black holes and weakly coupled gauge fields).

We are able to show that modular invariance of the CFT dual, together with compactness of the $U(1)$ gauge group, are enough to guarantee the existence of  light charged states, lighter than any charged black hole. We regard this as a version of the WGC in three dimensions. The recent work \cite{Benjamin:2016fhe} also uses modular invariance to place an upper bound on the conformal dimension of charged operators, in a more general setting than the one we consider (the results of \cite{Benjamin:2016fhe} are also valid for a noncompact gauge group, whereas ours  rely on compactness in an essential way).

There are however significant differences with the higher dimensional case: In three-dimensional black holes, electric charge is supported by a flat connection and thus does not backreact on the metric, even after higher derivative corrections are taken into account. As a result, electric charge thrown into a black hole resembles a global charge. This and other peculiarities of the three-dimensional case mean that the heuristic arguments based on remnants do not seem to have a place in three dimensions. Furthermore, black holes can also carry a discrete $\mathbb{Z}_N$ charge, which can be measured from infinity via an Aharonov-Bohm experiment. Unlike its integer-valued counterpart, modular invariance does not require the existence of light $\mathbb{Z}_N$ charged states, as we show in a specific example.

After this work was completed, we learned about \cite{HarvardPaper}, which applies the same spectral flow techniques to explore the WGC for perturbative closed string $U(1)$'s. Although the mathematical details and the spirit of the results are similar, the physical details (perturbative worldsheet vs. three-dimensional gravity theories) differ significantly.

The paper is organized as follows: Section \ref{sec:wgcrev} reviews the generic rationale for the WGC in higher dimensions. Section \ref{wgc3d} reviews the peculiarities of three-dimensional gravity and gauge fields. Section \ref{BHcft} discusses the relevant properties of the holographic dual. Section \ref{sec5} applies these properties to argue for the existence of light charged states, thus satisfying a form of the WGC. Section \ref{sec:zn} discusses the constraints that modular invariance poses on black holes with $\mathbb{Z}_N$ charge, for which the arguments of Section \ref{sec5} do not provide light charged states. Finally, Section \ref{conclus} contains our conclusions. Several important technical details have been relegated to the Appendices.

\section{The WGC in \texorpdfstring{$d>3$}{d=3}}\label{sec:wgcrev}
We will first revise the argument in \cite{ArkaniHamed:2006dz,Banks:2010zn} for the existence of light charged states in any weakly coupled $U(1)$+gravity theory. This will be convenient to highlight the differences to the three-dimensional case later on. Readers familiar with this argument may want to skip this Section.

Contrary to the three dimensional case, in $d>3$ gravity has local degrees of freedom. This means that one can build charged black hole solutions (the Reissner-Nordstron metric \cite{griffiths2009exact,Ortin:977337} and its higher-dimensional cousins, black branes of dimension $d-3$) and the metric is sensitive to the electric field of the black hole. When the tension  $M$ and charge $Q$ of the black object satisfy
\begin{align}M=\frac{g}{\sqrt{G}}Q\label{extr}\end{align}
the black hole is extremal, has zero temperature and is semiclassically stable. For $M<\frac{g}{\sqrt{G}}Q$, the solution presents a naked singularity, and thus it does not constitute a valid solution of the low-energy effective field theory \cite{Penrose:1969pc,Wald:1997wa}. As a result, there is a bound on the charge $Q$ that a black hole of given $M$ can stomach within the reach of effective field theory.  

The original WGC observation can be rephrased as the fact that, if we are allowed to tune gauge coupling to the limit $g\rightarrow0$, the maximum allowed charge  that a black hole of mass  $M$ can have blows up:
\begin{align}Q_{max}=\frac{M\sqrt{G}}{g}\rightarrow\infty.\label{asdhbbsa}\end{align}
 This means that, provided that $g\rightarrow0$ is a legitimate limit in the theory,  below any fixed mass $M_0$ there exists an arbitrarily large number of extremal black holes. If these objects are exactly stable, according to the standard lore, we will have a species problem and trouble with remnants \cite{Susskind:1995da,Banks:2006mm}. From the point of view of the low-energy effective field theory, each of these black holes will look like a particle of mass  $M\leq M_0$, and in a thermal bath at temperature $T$ (such as the one generically present whenever there is a horizon, for instance in the Unruh effect for a uniformly accelerated observer \cite{PhysRevD.14.870}) these black holes will contribute to the mean energy density as 
\begin{align} \rho(M,g) e^{-M/T} \label{thermalen}\end{align}
where $\rho(M)$ is the density of states at mass $M$. and coupling $g$.

From \eq{asdhbbsa}, we see that the integrated contribution of these objects diverges,  potentially rendering the theory pathological. At scales lower than  their mass, these states would renormalize the Planck constant. One-loop computations, as well as exact results in supersymmetric configurations \cite{Antoniadis:1997eg} suggest that an arbitrarily large amount of light particles would effectively turn off gravity in the deep IR. Alternatively, given a finite deep IR gravitational constant, the fast running induced by these states would make gravity strongly-coupled at a scale that goes to zero as $g\to 0$ (the cutoff $\Lambda\sim gM_P$ was suggested in~\cite{ArkaniHamed:2006dz}).

Thus, in a $d>3$ theory with finite Newton's constant in which naked singularities are forbidden, one of these two possibilities must take place:\begin{itemize}
\item It is not possible to take the $g\rightarrow0$ limit. In this case, we have nothing else to say, except that this does not seem to be the case in well known examples in string theory. 
\item The black holes do not give trouble of the kind suggested above because they are not stable. Unstable states only contribute significantly to a thermal bath for temperatures larger than their decay width $\Gamma$ (it is only then that the particle can exist in the thermal bath for a sufficiently long time). Semiclassically, an extremal black hole is exactly stable, but quantum-mechanically it may have a finite lifetime. These lifetimes can conspire to make the overall contribution \eq{thermalen} negligible. 
\end{itemize}

On what follows we focus on the second possibility. Generically, the black hole will decay by emitting some charged object. And, as stated above, it should be allowed to do so while remaining sub-extremal. Kinematically, this constrains the charge-to-mass ratio of the emitted object so that it is extremal or super-extremal \cite{ArkaniHamed:2006dz}
\begin{align}\label{eq:WGC} m\leq g \frac{q}{\sqrt{G}}.\end{align}
The existence of this object is (the electric form of) the WGC.  Both the WGC and its generalizations to several $U(1)$'s (such as those discussed in \cite{Cheung:2014vva,Brown:2015iha,Brown:2015lia,Heidenreich:2015nta}) or higher $p$-form gauge fields have always been validated in every string theory construction where it has been checked. Nevertheless, the arguments that led to~\eqref{eq:WGC} are not airtight and are based on several non-trivial assumptions, e.g. the Cosmic Censorship Hypothesis, the existence of the $g\rightarrow0$ limit, or the validity of the semiclassical description of black holes all the way to the Planck scale. Furthermore, as we will see, these arguments run very differently when applied to theories in three dimensions.  

\section{Gravity and gauge fields in three dimensions}\label{wgc3d}
There are qualitative differences in the physics of both gravity and matter between three and more spacetime dimensions. \cite{Deser:1983tn} gives a clear exposition of the peculiarities of gravity in the 3d case. Notably, there are no local degrees of freedom, and mass and angular momentum are given by topological invariants. A point mass only induces a conical defect in the geometry. As we review next, there are also peculiarities for $U(1)$ gauge theories.

\subsection{3D \texorpdfstring{$U(1)$ }{U(1)} gauge fields \& Chern-Simons terms}
In two space dimensions, the electric field sourced by a point charge falls of as $1/r$. As a result, the total electrostatic energy of a particle is IR divergent, as well as UV divergent. This already signals trouble with charged particles and gravity: The backreaction of a point particle on the geometry is significant, in the same way as for strings charged under a $B$-field in four dimensions or D7-branes in type II string theory \cite{Ibanez:2012zz,Denef:2008wq}. 

In the case of compact electrodynamics (gauge group $U(1)$ rather than $\mathbb{R}$, plus a single fermion) this  pathological IR behavior was understood long ago by Polyakov \cite{Polyakov:1976fu}. In three dimensions, the gauge coupling has dimensions\footnote{We are taking the convention with canonical kinetic term $-\frac12 F\wedge *F$.} of $[\text{length}]^{-1/2}$, so it shows classical running and in particular becomes strongly coupled in the IR. When there are monopoles in the theory, it shows confinement as measured by the  area behavior of the Wilson loop: the electrostatic energy between two particles is corrected from logarithmic to linear. As a result, no free charged particles are present in the theory (indeed, the low energy dynamics of the gauge field is that of a scalar field with a cosine potential).

There is another possibility which also solves the problem: In three dimensions,one can give a topological gauge-invariant mass to the $U(1)$ via a Chern-Simons term
\begin{align}\frac{N}{4\pi}e^2\int F\wedge A.\label{csterm}\end{align}
The low-energy dynamics again has a single degree of freedom (the transverse component of $A$), which propagates with mass $\mu=Ne^2/(2\pi)$. The monopoles (instantons in three dimensions) responsible for electric charge confinement in \cite{Polyakov:1976fu} now have no effect because gauge invariance forces them to spit charged particles; as a result, every allowed contribution to the vacuum path integral comes in monopole-antimonopole pairs, which give no potential overall: The Chern-Simons term \eq{csterm} confines monopoles, as opposed to electrically charged particles \cite{Affleck:1989qf}.

In this theory, electrically charged particles are allowed as predicted in the classical theory, but the electric field of a particle suffers a Yukawa-like screening, with no significant contributions at scales larger than $\sim 1/(Ne^2)$. This renders the electrostatic energy IR-finite. In what follows, we will focus on 3d $U(1)$ gauge theories with Chern-Simons term \eq{csterm}. As we will discuss in Section \ref{BHcft}, boundary terms in AdS require the Chern-Simons term to be present in order to have a unitary holographic dual.

\subsection{Consequences of compactness of the gauge group}
The Chern-Simons term has important consequences for the spectrum of the theory. The coefficient $N$ in \eq{csterm} must be an integer if the gauge group is compact. Compactness of the gauge group amounts to the identification of large Wilson lines
\begin{align}\int_{S^1} A=\frac{2\pi}{e}\ \sim \ \int_{S^1} A=0\label{gequiv}\end{align}
for any nontrivial $1$-cycle. This forces electric charges to be quantized in integer multiples of $e$.

The equation of motion
\begin{align}d*F=*j_e+\mu \label{eomg}\end{align}
can be integrated to yield a modified version of Gauss' law,
\begin{align}\int_{S^1}*F=Q_e+\mu \int_{S^1} A\end{align}
where $Q_e$ is the total electric charge enclosed by the $S^1$. As discussed above, one must have $F\rightarrow0$ faster than $r^{-1}$ at infinity for a configuration to have finite energy; as a result, the total electric charge satisfies
\begin{align}Q_e=-\frac{N}{2\pi}e^2\int_{S^1_{\infty}} A.\label{qishol}\end{align}
In other words, due to the Chern-Simons term, the electric charge sources a holonomy of the gauge field at infinity. This also implies that $N$ has to be an integer, at least for a compact gauge group, for a large gauge transformation changes $Q_e$ by $Ne$.

Although the long-range interaction is gone, one can still measure electric charge infinitely far away from a source by performing an Aharonov-Bohm experiment. However, due to \eq{gequiv}, only charge mod $N$ can be measured. This is a reflection of the discrete $\mathbb{Z}_N$ symmetry in the term \eq{csterm}; the Chern-Simons term is invariant not only under \eq{gequiv} but also under the shorter shifts
\begin{align}\int_{S^1} A\rightarrow \int_{S^1} A+\frac{1}{N}\frac{2\pi}{e}.\end{align}

While the Aharonov-Bohm phases are only senstive to a $\mathbb{Z}_N$ charge, this does not mean that electric charge is only conserved modulo $N$. In theories with Chern-Simons terms and monopoles, gauge invariance requires \cite{Blumenhagen:2006xt,Ibanez:2006da,Florea:2006si,BerasaluceGonzalez:2011wy} that any such monopole spits out $N$ units of electric charge. As a result, it seems that electric charge can be created or annihilated in multiples of $N$. However, the electric charge as measured at infinity via \eq{qishol} does not change; the contribution from the newly created electric particles is cancelled by that of the field of the monopole. Electric charge is exactly conserved, as can be proven straightforwardly in the dual CFT; this fact will play a significant role in Section \ref{sec:specflow}.

\subsection{Black hole solutions}\label{blackholesols}
Although the issues with the IR behavior of the $U(1)$ have been resolved satisfactorily, the flat space theory contains no black holes. This is not the case in AdS space, in which case we have the famous BTZ black holes \cite{Banados:1992wn}, which we now briefly review. For further reference, check \cite{Carlip:1995qv,Birmingham:2001dt,Carlip:2005zn,Kraus:2006wn}.

The BTZ black hole with mass $M$ and angular momentum $J$ is described by the metric 
\begin{align}ds^2&=-N^2(r)dt^2+\frac{dr^2}{N^2(r)}+r^2\left(d\phi+N^\phi(r)dt\right)^2,\nonumber\\N(r)&=-8GM+\frac{r^2}{l^2}+16\frac{(GJ)^2}{r^2},\quad N^{\phi}(r)=-4\frac{GJ}{r^2} dt.\label{btzmetric}\end{align}

This metric has a horizon at 
\begin{align}r_+=l\left[4GM\left(1+\sqrt{1-\left(\frac{J}{Ml}\right)^2}\right)\right]^\frac{1}{2},\label{rhor}\end{align}
which exists as long as $\vert J\vert\leq Ml$. $M$ is the ADM  mass and $J$ is the angular momentum at infinity. 

This black hole has been studied extensively in the literature, and shares many properties with its higher-dimensional cousins. Namely, it has a finite horizon area, Hawking radiation \cite{Hyun:1994na}, and arises as the collapse of dust in an AdS background \cite{Ross:1992ba}.  

 The charged BTZ black hole  in theories with Chern-Simons terms \eq{csterm} has been studied in \cite{Andrade:2005ur}. The presence of a horizon is only compatible with no electric field emanating from the black hole.  This is reminiscent of the standard no-hair theorems: Much like the 4d Schwarzschild solution, the BTZ black hole satisfies a scalar no-hair theorem \cite{Birmingham:2001dt}. A $U(1)$ with a term \eq{csterm} is similar to a massive scalar, as discussed above, and thus no scalar hair is the same (at least as far as equations of motion are concerned) as no gauge hair. 

This, however, does not mean that there are no charged black holes. The BTZ geometry \eq{btzmetric} has a non-contractible 1-cycle which can support a flat connection for the gauge field. As discussed around \eq{qishol}, such a connection measures the electric charge of the black hole. Thus, there are indeed charged BTZ black holes, but classically they are identical in every way to the ordinary BTZ black hole except for an extra flat connection on the non-contractible $S^1$. One can measure the $\mathbb{Z}_N$ electric charge of the black hole from infinity via an Aharonov-Bohm effect\footnote{ Much like in~\cite{Coleman:1991ku}, there is an exponentially vanishing electric field around the black hole from which the discrete $\mathbb{Z}_N$ charge can also be measured. Alternatively, the Hawking temperature will also be affected and provides yet another way of measuring the $\mathbb{Z}_N$ charge.}.

To sum up, existence of black hole solutions requires the spacetime to be AdS, and reasonable IR $U(1)$ interactions can be obtained via a Chern-Simons term \eq{csterm}, which as we will see in Section \ref{BHcft} it is also strongly suggested by holographic considerations. These conditions imply two major differences with the situation in higher dimensions, which prevent us from emulating the too-many-remnants argument for the WGC in Section \ref{sec:wgcrev}:\begin{itemize}
\item Because only a $\mathbb{Z}_N$ charge is observable in the deep IR, there is no dependence on the gauge coupling and no meaningful way in which we can take the limit $e\rightarrow0$ far away from the black hole.
\item The charged BTZ solution does not have an obvious notion of extremality (other than the $J=Ml$ limit),
\end{itemize}
The first point can be understood as a result of the dimensionful character of the gauge coupling; in the IR, formally the coupling goes to infinity and thus it is not consistent to take the $e\rightarrow0$ limit in the IR; it is always strongly coupled there. The Chern-Simons terms introduces a mass gap so that the deep IR theory should be trivial. We can only give a proper understanding of the second point in a holographic theory, which we will do in section \ref{BHcft}.

\subsection{Black hole discharge in AdS}

Before discussing the problem from a holographic perspective, a brief discussion of the kinematical peculiarities of AdS relevant to the WGC is in order.

In AdS space, black hole solutions fall into two categories \cite{Hawking:1982dh,Myung:2006sq,Eune:2013qs,Hartman-lect}: Large AdS black holes, which are bigger than the AdS radius, and small black holes. The former are in equilibrium with their own thermal bath, and are thus stable. If we consider small black holes instead, any WGC should demand the existence of a super-extremal state in the theory for each value of the charge, $Q$. 

This is in contrast with the situation in Minkowski space, where one can obtain super-extremal black holes for each value of $Q$ just by having a single super-extremal species $p$ of mass $m$ and charge $q$. A black hole of large charge $Q$ can radiate away all of its charge to infinity in the form of a cloud of $Q/q$ non-interacting $p$'s.  As these particles go further away from the black hole they become more and more weakly coupled, and in the limit of infinite separation they constitute a super-extremal state of charge $Q$. 

However, the usual AdS space boundary conditions behave like a box, in the sense that signals emitted from the bulk are reflected back and never reach the boundary. As a result, any particles emitted by an AdS black hole, evaporating or not, will eventually interact strongly. As a consequence, mild versions of the WGC, which only require the existence of a single super-extremal particle $p$, are not enough to ensure black hole decay in AdS spaces: The cloud of emitted particles will eventually bounce back and start interacting; there is no guarantee that the interactions do not render the condensate sub-extremal.

As a result, in AdS a Strong/Lattice version of the WGC is the most natural one: In each sector of charge $Q$, there should be a super-extremal state. While in some cases these states might be described as strongly interacting condensates of a single particle, in the spirit of the mild form, this description is not necessary (or generically very useful).

\section{3d black holes from the CFT perspective}\label{BHcft}
So far we have made general considerations which apply to any weakly coupled theory containing gravity and a $U(1)$ gauge symmetry. To be able to say more we need more control and thus from here on we will assume the existence of a holographic dual. This will allow us to define a black hole threshold (see e.g. \cite{Benjamin:2016fhe}) for operators corresponding charged BTZ black holes in subsection~\ref{sec:ebound}, which is the CFT version of an extremality bound.

 We will begin with a brief summary of the properties of the $AdS_3/CFT_2$ correspondence which will be of relevance to our study. The reader is referred to \cite{Kraus:2006wn} for further details.

\subsection{Extended chiral algebra and the Sugawara construction}
We will consider a $2+1$ theory with a weakly coupled limit including gravity and a $U(1)$ gauge field, while also assuming the existence of a holographic CFT dual. In this case, the central charge of the CFT is related to the AdS radius $l$ (which itself is related to the cosmological constant as $-\Lambda=l^{-2}$) and Newton's constant in 3d by
\begin{align}c=\frac{3l}{2G}\label{ccharge}.\end{align}
The semiclassical treatment is only valid for small curvature, that is, large central charge. 

 The gauge $U(1)$ symmetry of the bulk theory is translated to a global $U(1)$ symmetry in the CFT language: this is precisely the global part of the bulk gauge group. This is so because in low-energy EFT, a bulk $U(1)$ must couple to a conserved current: its boundary value provides a conserved current in the CFT (a primary operator of conformal weights $(1,0)$ or $(0,1)$) which generates this symmetry.
 This is described in more detail in appendix \ref{dualcurrent}. Here we just summarize the results: A gauge bulk theory with a Chern-Simons level $N>0$ induces a holomorphic current $j(z)$ in the boundary CFT. This current extends the chiral symmetry algebra of the theory, supplementing the Virasoro algebra with extra generators $j_m$ (the Laurent coefficients in $j(z)=\sum_n j_m z^{-(m+1)}$) which satisfy \cite{Polchinskiv2,Goddard:1988fw}
\begin{align}[j_m,j_p]=N\delta_{m+n,0},\quad [L_m,j_p]=-pj_{p+m}.\label{eca}\end{align}
For this reason theories with additional conserved currents are said to have extended chiral algebras. Of particular importance is the generator $j_0$, which is the center of the extended chiral algebra; unitary representations will be labeled by the eigenvalues of $j_0$, in addition to the highest weight $h$ of the representation. As shown in appendix \ref{dualcurrent}, the eigenvalue of $j_0$ is precisely \eq{qishol}, the electric charge under the bulk $U(1)$ of the configuration, in multiples of the fundamental electric charge $e$. 

For $N=0$, the algebra \eq{eca} does not admit nontrivial unitary representations\footnote{ When $N=0$, the set $\{j_m\}$ can be simultaneously diagonalized together with $L_0$, so that $j(z)\ket{0}=f(z)\ket{0}$ for some holomorphic function $f(z)$, as long as the vacuum state $\ket{0}$ is unique. By the state-operator mapping, in a unitary theory this implies that $j(z)=f(z)\mathbf{I}$, where $\mathbf{I}$ is the identity. This is an operator of weight $(0,0)$ instead of the expected $(1,0)$, unless we are in the degenerate case $f(z)=0$.}. Thus, as advertised in the previous Section, consistency of the holographic dual of a three-dimensional $U(1)$ gauge field requires $N>0$, i.e. the presence of Chern-Simons terms in the bulk action.

The CFT evolution operator is $L_0+\tilde{L}_0$, which according to \eq{eca} commutes with the electric charge. Therefore, electric charge, and not just electric charge mod $N$, is an exactly conserved quantity in the theory. 

An important property of theories with extended chiral algebras is the Sugawara construction \cite{Polchinskiv2,Goddard:1988fw}. The stress-energy tensor $T(z)$ of the theory decomposes as
\begin{align}T(z)=T'(z)+T^S(z),\quad T^S(z)=\frac12 :j(0) j(z):,\label{sugawara}\end{align}
where $T'(z)$ has vanishing OPE with the currents. The Virasoro generators similarly split as $L_m=L_m'+L_m^S$, where the Sugawara generators $L_m^S$ satisfy a Virasoro algebra with central charge $c^S$ equal to the rank of the bulk gauge group (if abelian). The primed generators $L_m'$ satisfy an independent Virasoro algebra with central charge $c-c^S$, where $c$ is the central charge of the full Virasoro generators $L_m$. Clearly, $c>c^S$ in a unitary theory. This fact has a nice interpretation in terms of the dual AdS: The central charge essentially measures the strength of gravity, which becomes weaker as we increase $c$. $c^S$ is the number of massless $U(1)$ gauge fields present in the theory.  Thus, $c>c^S$ is just the familiar statement that massless fields tend to renormalize Newton's constant, making gravity weaker. 

An important constraint comes from the split $L_0=L'_0+L_0^S$. In a unitary representation of the conformal algebra, $L_0$ is positive definite \cite{Polchinskiv2,Goddard:1988fw}, and so is $L_0'$. Therefore, $\langle L_0\rangle\geq \langle L_0^S\rangle$. This last operator in turn satisfies 
\begin{align}L_0^S\geq \frac{Q^2}{2N},\quad j_0\ket{\psi}=Q\ket{\psi}.\label{unit-bound}\end{align}
This is a unitarity bound which shows that the eigenvalues of $L_0$ cannot become arbitrarily small for a given charge.

The conformal dimension $\Delta$ of an operator of weights $(h,\tilde{h})$ is simply $\Delta= h+\tilde{h}-\frac{c+\tilde{c}}{24}$. We will denote by $\Delta_0=-\frac{c+\tilde{c}}{24}$ the conformal dimension of the vacuum. The AdS/CFT correspondence generically relates the conformal dimension $\Delta$ to the energy of the bulk configuration. As a particular simple example, a scalar operator of dimension $\Delta$ corresponds to a scalar field in AdS with a mass given by \cite{Witten:1998qj}
 \begin{align} m^2l^2=(\Delta-\Delta_0)(\Delta-\Delta_0-2)\label{ssa}\end{align}
 Thus, for large dimension, $m\approx \Delta/l$. The unitarity bound \eq{unit-bound} corresponds to 
 \begin{align}m\gtrsim \frac{Q^2}{2N l},\label{uboundads}\end{align}
 so that electric charge necessarily gravitates; we cannot pile up an arbitrarily large amount of charge without the corresponding increase in energy.
 
 \subsection{An extremality bound}\label{sec:ebound}
The above considerations impact the computation of black hole entropy via evaluation of the bulk action in a significant way. The Sugawara contribution to the stress-energy tensor is also visible in the bulk. As discussed in appendix \ref{dualcurrent}, it is necessary to supplement the bulk action with a boundary term proportional to $\int_{\partial AdS} A\wedge *A$. This term depends manifestly on the metric via the Hodge dual, so that it gravitates, and it can be shown (see, e.g., \cite{Kraus:2006wn,Jensen:2010em}) that its contribution precisely agrees with \eq{sugawara}. 

Specifically, there is a boundary contribution to the stress-energy tensor,
\begin{align}T^S_{ij}=\frac{N}{4\pi}\left( A_i A_j-\frac12 A_k A^k g_{ij}\right).\end{align}
in addition to the to the Brown-Henneaux stress-energy tensor, which measures the contribution of the geometry,
\begin{align}T'_{ij}=\frac{1}{8\pi G l}\left(g_{ij}^{(2)}-\text{Tr}(g^{(2)})g_{ij}^{(0)}\right),\label{bhenneaux}\end{align} 
where $g^{(0}_{ij}$ represents the ``conformal boundary metric'' and $g^{(2)}_{ij}$ its first correction in an asymptotic expansion of the metric (see, e.g.,~\cite{Kraus:2006wn}).
This decomposition of the total stress-energy tensor yields a similar decomposition of its Fourier modes, which are precisely the Virasoro generators. In particular, we recover the Sugawara decomposition of the $L_0$ generator, $L_0=L_0^S+L_0'$. 

In a spacetime containing a BTZ black hole of mass $M$ and angular momentum $J$, we can compute the $L_0'$ contribution exactly, from \eq{bhenneaux}. The eigenvalues $h'_{M,J},\tilde{h}'_{M,J}$ of the gravitational part of the stress energy tensor are \cite{Kraus:2006wn} 
\begin{align}h'_{M,J}=\frac{c}{24}+\frac12(Ml+J),\quad \tilde{h}'_{M,J}=\frac{\tilde{c}}{24}+\frac12(Ml-J).\label{weightBTZ}\end{align}

As a result, we can assign a weight to the microstates comprising a black hole of mass $M$, charge $Q$, and angular momentum $J$, as
\begin{align}h_{M,J}=\frac{c}{24}+\frac12(Ml+J)+\frac{Q^2}{2N},\quad \tilde{h}_{M,J}=\frac{\tilde{c}}{24}+\frac12(Ml-J).\label{weightBTZ2}\end{align}
Equation \ref{weightBTZ2} has the following interpretation from the CFT perspective: In the high temperature limit, the partition function (in the sector of charge $Q$) is dominated by a (charged) black hole. This means that an overwhelming majority of the CFT states contributing to the partition function have a semiclassical description in the bulk containing a macroscopic BTZ black hole. These majority of states have a weight given by \eq{weightBTZ}, to a very good approximation. 

In other words, only operators above the black hole threshold
\begin{align}h \gtrsim h_{M,J}=\frac{c}{24}+\frac12(Ml+J)+\frac{Q^2}{2N} ,\quad \tilde{h}\gtrsim \tilde{h}_{M,J}=\frac{\tilde{c}}{24}+\frac12(Ml-J)\label{exboundcft}\end{align}
can be dual to a bulk configuration with a semiclassical charged BTZ black hole of mass $M$ in it. States satisfying the bound\eq{exboundcft} typically have a semiclassical description containing a macroscopic, charged black hole. Thus, \eq{exboundcft} is the CFT version of an extremality bound. Any operator significantly below the bound \eq{exboundcft} will correspond, in the bulk, to a charged state lighter than the lightest black hole; a super-extremal object, by any account. 

From here on, we will refer to~\eqref{exboundcft} as the three dimensional extremality bound. States associated to operators that satisfy ~\eqref{exboundcft} will be referred to as sub-extremal (extremal if the bound is saturated), and states that violate it will be denoted super-extremal. We say that the WGC is satisfied in three dimensions in a sector of charge $Q$ if there exists a super-extremal state with such a charge. We will also sometimes refer to those states as ``WGC-states''. Sections~\ref{sec5} and~\ref{sec:zn} explore the existence of such states in different charge sectors.

Equation \eq{exboundcft} refers to the case of a single left-moving current; the parity symmetric case should have the corresponding Sugawara term in $\tilde{h}_{M,J}$. For the parity-symmetric case, one can take $h=\tilde{h}$ and thus $J=0$. \eq{exboundcft} now becomes a bound on the conformal dimension of the operators, 
\begin{align}h+\tilde{h}\gtrsim Ml+\frac{c}{12}+\frac{Q^2}{N}.\label{exj0}\end{align}
For $Q=0$, we recover the familiar black hole threshold \cite{Banados:1992wn, Benjamin:2016fhe}; BTZ microstates correspond to operators of large dimension.

\subsubsection*{The entropy bound from the bulk and boundary perspectives}

Unlike the higher-dimensional case, the extremality bound \eq{exboundcft} does not show up immediately as a consequence of the backreaction of the gauge field on the geometry. On the contrary, the charge is carried by a flat connection, which is locally pure gauge and therefore cannot backreact classically, even after higher derivative corrections are taken into account.

To further motivate \eq{exboundcft} as the relevant extremality condition, we will now show that it is also in agreement with the computation of black hole entropy both from CFT and AdS points of view. For simplicity, we will work in the parity-symmetric, $J=0$ case; the task is then to recover \eq{exj0}. 

From the CFT side, we need the partition function of the CFT in the sector of charge $Q$ in the high temperature $\beta\rightarrow0$ limit. As shown in Appendix \ref{app:cardy}, due to modular invariance this is
\begin{align}Z_Q(\beta)\approx e^{-\frac{4\pi^2}{\beta}\Delta_0-\frac{Q^2}{N}\beta},\label{cformain}\end{align}
a slightly modified version of Cardy's formula. The thermodynamic energy ($\langle h+\tilde{h}\rangle-\frac{c}{12})$ is 
\begin{align}\langle h+\tilde{h}\rangle-\frac{c}{12}=-\frac{\partial \log Z_Q}{\beta}=-\frac{4\pi^2}{\beta^2}\Delta_0+\frac{Q^2}{N}\end{align}
from which we obtain an entropy
 \begin{align}\frac{S}{4\pi}=\sqrt{\frac{c}{12}\left( \langle h+\tilde{h}\rangle -\frac{c}{12}-\frac{Q^2}{N}\right)}\label{BHent}.\end{align}

This again reproduces the black hole threshold; We need operators of dimension higher than $c/12+Q^2/N$ if we want to be in the regime where \eq{BHent} is trustworthy. On the other hand, from the bulk perspective the entropy is obtained by evaluating the action of the Euclidean spacetime. In a BTZ spacetime, we will have the usual BTZ black hole entropy, which depends on the mass $M$ and comes from the boundary terms of the gravitational action, plus the Sugawara term, which is simply $-\frac{Q^2}{N}\beta$. As a result, the bulk entropy is
\begin{align}S=\frac{2\pi r_+}{4G}=\frac{2\pi l}{4G}\sqrt{8 GM}\label{bulkentr}\end{align}
Equality of \eq{BHent} and \eq{bulkentr}, using \eq{ccharge}, imposes again \eq{weightBTZ2}.
 
 The fact that the extremality bound depends on the total electric charge, and not only on the charge modulo $N$, makes it clear that $Q$ is exactly conserved, even if only its $\mathbb{Z}_N$ part can be measured via an Aharonov-Bohm experiment. We will have more to say about the extremality bound in the next Section.
 
 \section{Super-extremal states in \texorpdfstring{$AdS_3$}{AdS3}}\label{sec5}

 \subsection{Modular invariance}\label{modinv}
 On top of the ingredients described in the previous Section, we should require the CFT to be modular invariant. This is often taken as a key requirement for CFT's dual to gravity theories. As explained in \cite{Dijkgraaf:2000fq,Keller:2014xba} (see \cite{Kraus:2006wn} for a review), modular invariance can be understood from the fact that the AdS/CFT prescription instructs us to sum over every geometry with the same asymptotic behavior. The CFT partition function on the torus is dual to a sum over every geometry with flat AdS asymptotics. This turns out to include black hole states in a $SL(2,\mathbb{Z})$ multiplet in such a way that the resulting partition function is modular invariant. Thus, modular invariance is related to the existence of black hole states in AdS. In fact, we expect the high temperature partition function to be dominated by a single black hole semiclassical contribution, and thus the high-temperature entropy of the CFT should reproduce \eq{BHent}. This is famously the case if we apply Cardy's formula, which requires modular invariance.

Modular invariance is a strong constraint on a theory: It forces a GSO projection, the equality of the left and right-moving central charges (modulo an integer), and for theories with extended chiral algebras in which the $U(1)$ sector is modular invariant by itself it forces the charge lattice (the $n$-dimensional lattice spanned by the charge vectors $j_0^a$ of every operator in the theory) to be even and self-dual \cite{Polchinskiv2,Ibanez:2012zz}.

We should stress that throughout this paper we require modular invariance of the statistical partition function, defined as a trace $\text{Tr}(\exp^{-\beta H})$ over all the states of the theory, possibly with extra insertions to accomodate a chemical potential. This is unlike the worldsheet CFT partition functions considered in \cite{HarvardPaper}, which lack a direct statistical interpretation and can vanish, typically in supersymmetric configurations. 

We will see, in the remainder of this Section, how modular invariance readily implies a statement along the lines of the Weak Gravity Conjecture.

 \subsection{The spectral flow \& super-extremal states}\label{sec:specflow}
 As discussed in Section \ref{blackholesols}, the effective field theory in AdS retains invariance under global large gauge transformations, and in particular under shifts of the flat connection of the black hole $Q\rightarrow Q+nN$, for any integer $n$. This property, which is only present for theories with a compact gauge group, will be essential in establishing the WGC in three dimensions. 
 
We will now describe how invariance under large gauge transformations of the charged BTZ black hole is implemented in the CFT. The extended chiral algebra \eq{eca} admits a nontrivial automorphism \cite{Goddard:1988fw, Polchinskiv2, Kraus:2006wn}
\begin{align} L_n\rightarrow L_n+\mu j_n + N\mu^2,\quad j_n\rightarrow j_n+N\mu\delta_{n,0},\quad \mu\in\mathbb{R}\label{specflow}\end{align}
known as spectral flow. To our knowledge, this automorphism has been mostly used in supersymmetric setups to constrain the elliptic genus \cite{Schwimmer:1986mf,Manschot:2008zb,Gaberdiel:2008xb,VanHerck:2009ww,Hellerman:2010qd,Keller:2013qqa}. Physically, it shifts the electric charge $Q$ (measured by $j_0$) by $\mu$. Similar considerations would apply to antiholomorphic currents, putting a tilde on top of everything. 

Spectral flow by $\mu$ is not a symmetry of the theory for generic $\mu$. However, acting on some operator operator of charge $Q$ whose bulk description resembles a BTZ black hole, the transformation \eq{specflow} turns it into another black hole state of charge $Q+N\mu$. Hence, for $\mu=n$, the spectral flow automorphism maps microstates of a black hole of mass $M$ and charge $Q$ to microstates of a black hole of charge $Q+nN$ and mass $M+N/l$, in order to comply with the extremality bound \eq{exboundcft}. In the highly subextremal limit, $M\gg Q^2/(Nl)$, we can neglect the change in mass; the effect of the spectral flow is just to shift the charge, by an integer multiple of $N$, precisely in agreement with the behavior of the large gauge transformation of the semiclassical theory.

 Spectral flow by $N$ units leaves the spectrum of large black hole states invariant, but modular invariance relates these high temperature solutions to the AdS vacuum. As a consequence, spectral flow by $N$ units becomes an exact symmetry of the full theory. We prove this, directly in the CFT, in Appendix \ref{app:specflow}, and now discuss the interesting consequences. In particular, acting with the spectral flow on the identity operator yields a state of charge $N$ and (in compliance with \eq{unit-bound}) $h=N/2,~ \tilde{h}=0$, i.e.
\begin{align}\Delta-\Delta_0=\frac{N}{2},\quad J=\frac{N}{2},\quad Q=N.\label{wgc-specflow}\end{align}

In this way, modular invariance requires the presence of light charged (super-extremal) operators, and hence the existence of bulk states that fulfill the WGC in the sector of charge $Q=N$.  In fact, these states are as light as they can be, since they saturate \eq{unit-bound}. As a consequence, the operators satisfying \eq{wgc-specflow} are primaries; their bulk interpretation is a massive state sitting in the center of AdS \cite{Hellerman:2009bu}. In the parity symmetric case, where this state is a scalar, the AdS mass of the state is given by \eq{ssa}. This argument, applied to the worldsheet CFT, was used in \cite{ArkaniHamed:2006dz} to prove the Weak Gravity Conjecture in perturbative heterotic CFT models. See also \cite{HarvardPaper} for a more in-depth discussion of the subtleties of this approach.

For level $N=1$, equation \eq{wgc-specflow} qualifies as an example of a Lattice Weak Gravity Conjecture: For each $Q$, it predicts the existence of a (very) super-extremal state, since we can apply the spectral flow automorphism as many times as we want. For $N=1$, the spectral flow states of \eq{wgc-specflow} and their Virasoro descendants allow small charged black holes to decay to radiation. If no charged states below \eq{exboundcft} existed, then a small charged black hole would not be able to evaporate completely, resulting in a remnant.  However, this is not as troublesome as in higher dimensions, since the too-many-remnants arguments of Section \ref{sec:wgcrev} do not apply: The gauge coupling is dimensionful, and it only depends on the AdS radius and the current algebra level (which we are setting to one for now). As a result, there is no parameter for us to tune, and no troublesome limit in which we get too many mass degenerate black hole microstates. 

Perhaps one may argue instead that in three dimensions the WGC states exist in order to accommodate a Hawking-Page transition in the charged sectors : At large temperature, the partition function in the sector of charge $Q$ is dominated by a large black hole. If there is a Hawking-Page phase transition at some temperature, the partition function at low temperature should be dominated by radiation over AdS. This will only be the case if there is a light enough excitation over AdS. This argument is even less compelling than the ones based on remnants in Section \ref{sec:wgcrev}: We do not see anything evidently wrong with a theory in which the lightest charged state is above the black hole threshold and there is no Hawking-Page transition. In any case, since the Hawking-Page phase transition of BTZ black holes is also related to modular invariance \cite{Hartman-lect,Eune:2013qs}, this does not seem an unreasonable connection.

For $N>1$, \eq{wgc-specflow} only provides super-extremal states with charge multiple of $N$, in other words, only for sectors of vanishing $\mathbb{Z}_N$ charge (the measurable Aharonov-Bohm phase of the black hole). Therefore,  spectral flow arguments are not enough to show that a version of the WGC holds for the discrete $\mathbb{Z}_N$ charge. Although the general remnant arguments for the WGC do not hold anyway for discrete symmetries, Section \ref{sec:zn} studies what can be said in this case.

Notice that the operator predicted by \eq{wgc-specflow} has nonvanishing angular momentum, and in fact it has the maximal $J$ compatible with the unitarity constraint $\tilde{h}>0$\footnote{Such states are often called extremal. The BTZ metric \eq{btzmetric} with $J=Ml$ is on the verge of developing a naked singularity. We will avoid this terminology in order to avoid a clash with our charge-dependent notion of extremality \eq{exboundcft}}. This is related to our assumption of a holomorphic current in the CFT (equivalently, a  $U(1)$ with positive Chern-Simons level). Generally, one may consider an extended chiral algebra with both holomorphic and antiholomorphic currents. Spectral flow under a linear combination of these currents shifts $h$ and $\tilde{h}$ accordingly. Of particular interest is the case of parity symmetric theories. In AdS, parity takes a $U(1)$ with level $N$ Chern-Simons to another one with level $-N$, and thus in the CFT requires the presence of both left and right-moving currents at level $N$. One may consider spectral flow along the diagonal combination of both $U(1)$'s, which yields 
\begin{align}\Delta-\Delta_0=N,\quad J=0,\quad Q=N\label{wgc-sp-diag}.\end{align}
In AdS, this corresponds to a scalar field of mass $\sim N/l$. Notice that while in the purely holomorphic case \eq{wgc-specflow} the condition of modular invariance $h-\tilde{h}\in\mathbb{Z}$ forces us to consider even $N$, no such restriction exists for the parity-symmetric case.

The spectral flow argument is easily generalized to theories with more than one $U(1)$. In this case, the current algebra \eq{eca} is modified to
\begin{align}[j^a_m,j^b_n]=N^{ab}\delta_{0,m+n},\end{align}
the Sugawara term is now
\begin{align}L_0^S=\frac12N^{-1}_{ab}Q^aQ^b\label{sug2}\end{align}
and the spectral flow provides light charged states only for a \emph{sublattice} of the charge lattice (spanned by  the columns of $N^{ab}$, where $v_b$ is a generic vector in the dual of the charge lattice). Thus, the spectral flow argument only supports a Sublattice WGC instead of the full Lattice WGC introduced in \cite{Heidenreich:2015nta}. A sublattice WGC is in many ways more similar to a mild form of the WGC than to a strong form. For instance, the sublattice WGC would allow for the loophole discussed in \cite{Rudelius:2015xta,Montero:2015ofa,Brown:2015iha,Bachlechner:2015qja,Hebecker:2015rya,Brown:2015lia,Palti:2015xra} to argue for transplanckian field ranges in natural inflation.

Recently \cite{Benjamin:2016fhe} has obtained an upper bound on the conformal dimension of the lightest charged state, in terms of the weight of the vacuum $\Delta_0=-\frac{c+\tilde{c}}{24}$,
\begin{align}\Delta-\Delta_0<\frac{c}{6}+\frac{3}{2\pi}+\mathcal{O}(c^{-1})\label{wgc-mod-bootstrap}\end{align}
which improves to $\Delta<1+\mathcal{O}(c^{-1})$ for $(1,1)$ supersymmetric theories. These bounds are more general than \eq{wgc-specflow}, as we have assumed compactness of the gauge group, whereas no such assumption has been made in \eq{wgc-mod-bootstrap}. For the compact case, the level of the current algebra has a physical meaning; it provides a metric on the charge lattice, via the Sugawara construction. Although it is always possible to rescale the current so that $N=1$, the charge lattice changes in the process, and different lattices correspond to different Dirac quantization conditions. On the other hand, \cite{Benjamin:2016fhe} shows the existence of a single light charged operator, whereas as we argued in Section \ref{blackholesols} we would expect to have one for each charged sector of the theory. Furthermore, \eq{wgc-mod-bootstrap} does not give in general a state below the black hole threshold \eq{exboundcft}. The power of \eq{wgc-mod-bootstrap} however lies in its generality, as it can be applied to any situation with an abelian current algebra.

\subsection{3d black holes and global symmetries.}\label{qgglob}

Not only the usual WGC arguments based on remnants do not  seem to hold in the three-dimensional case; even the usual argument for the absence of global symmetries in quantum gravity does not directly apply to BTZ black holes. The usual wisdom is that a black hole charged under a global symmetry does not have any hair giving it away, and thus black hole evaporation leaves us with potentially infinitely many different microstates. Continuous gauge symmetries escape this argument because the charge can be measured at infinity via Gauss' law, and the resulting solutions come with an extremality bound which forbids large pile-ups of charge without a corresponding increase in the mass.

Although the $U(1)$ charge discussed throughout this paper is gauge, it behaves more like a global charge when thrown into a BTZ black hole. For $N=1$, the Aharonov-Bohm phase of any charged particle circling the black hole is trivial, and the external electric field also vanishes.  There seems to be no way to distinguish from the outside two BTZ black holes with different values of electric charge $Q$.

Furthermore, Hawking radiation is insensitive to $Q$. From the point of view of the low-energy effective field theory discussed in Section \ref{wgc3d}, the two black holes are related by a large gauge transformation of the form
 \begin{align}A\rightarrow A+\frac{d\phi}{e}\label{saswqe}\end{align}
  i.e. by a Wilson line in $H_1(X,\mathbb{Z})$, supported on the nontrivial 1-cycle of the BTZ spacetime. Upon this transformation the action shifts by the boundary term \eq{bterm}, so it is a symmetry of the classical theory. This is reflected on the entropy formulas of Section \ref{sec:ebound}, which only depend on the black hole mass; as a result, Hawking radiation is independent of $Q$. Alternatively, we can look at the original computation of Hawking radiation, which is only sensitive to the equation of motion and the inner product chosen on the space of solutions   (see \cite{Hyun:1994na} for the relevant BTZ black hole case). Coupling to a flat connection only modifies the equation of motion of that reference by the replacement $\nabla_\mu\rightarrow\nabla_\mu-iqg A_\mu$. If we consider a charged BTZ black hole, $A_\mu$ is just a flat connection $A=Qd\phi$, which can be reabsorbed via a gauge transformation acting on the charged field $\psi$ as $\psi\rightarrow e^{i q Q\phi}\psi$. As a result, we end up with the same equation of motion as for $A_\mu=0$, but the angular momentum of each solution is shifted by $qQ$. The set of all solutions, however, and the Bogoliubov coefficients (which give directly Hawking temperature) remain the same. Finally, the computation of the Hawking temperature in the Euclidean formalism (where it is chosen to avoid a conical singularity in the Euclidean solution) is again manifestly independent of the black hole charge.

 So, for all effects and purposes in the semiclassical theory, the electric charge of black holes behaves like a global charge. However, the usual way to get into trouble with remnants (building a very large black hole and then waiting for it to evaporate via Hawking radiation) does not work, as a BTZ black hole larger than the AdS radius will not evaporate. Even if we were somehow able to put an arbitrarily large amount of charge in a black hole of radius $r\leq l$, such a black hole has a mass lower than the three-dimensional Planck mass. As a result,  there will be generically large quantum corrections which make the semiclassical picture of Hawking radiation and even the BTZ metric itself unreliable. 

   To sum up, on one, hand, large black holes in AdS do not decay and do not lead to a large number of light remnants, whereas  in three dimensions quantum effects are important even for black holes much larger than the Planck length. As a result, there is no apparent contradiction with the semiclassical unobservability of the black hole electric charge, in contrast with the situation in $d>3$.
   
From a holographic point of view, a gauged symmetry in the bulk is in a one to one correspondence with a conserved current in the CFT. Thus, in AdS/CFT the statement that symmetries are gauged in quantum gravity is equivalent to Noether's theorem in the CFT \cite{Beem:2014zpa}. While this is not a proof that all continuous symmetries are gauged, given the fact that it is not clear that Noether's theorem holds for nonlagrangian theories, it is a more compelling argument than those based in remnants.

\section{Constraints on theories with level \texorpdfstring{$N$}{N} current algebras}\label{sec:zn}
Although the results of Section 4 constitute a version of the WGC for a $U(1)$ in three dimensions, it only predicts operators with charge an integer multiple of $N$. In this case there is a $\mathbb{Z}_N$ charge, given by the original $U(1)$ charge mod $N$, for which spectral flow provides no light states. In other words, for a theory with a level $N$ Chern-Simons term black holes can have an extra $\mathbb{Z}_N$ charge, which is measurable at infinity via an Aharonov-Bohm experiment. 

In this Section, we will find out just how much can modular invariance alone tell us about this $\mathbb{Z}_N$ charge.

\subsection{Modular invariance constraints}\label{modznconsts}
As mentioned above, the spectral flow automorphism implies invariance of the spectrum under a shift of $N$ units of charge. The theory splits into $N$ sectors with different values of charge modulo $N$. The partition functions in these sectors are also heavily constrained by modular invariance, as we will now see. We will start with the usual partition function with chemical potential,
\begin{align}Z(z,\tau)= \sum \exp(-2\pi i Q_0 z) q^{L_0-\frac{c}{24}}\bar{q}^{\bar{L}_0-\frac{\tilde{c}}{24}}.\end{align}
The partition function in the sector of $\mathbb{Z}_N$ charge $Q_0$ (that is, charge $Q_0$ mod $N$) is related to the above via
\begin{align}Z_{Q_0}(\tau)=\frac{1}{N}\sum_{k=0}^{N-1} e^{2\pi i Q_0 \frac{k}{N}} Z(\frac{k}{N}, \tau).\label{pfuncch}\end{align}

 We may regard $Z_{Q_0}$ as a function on $\mathbb{Z}_N$. Eq. \eq{pfuncch} tells us that the discrete Fourier transform of $Z(z,\tau)$ is precisely $Z_{Q_0}$. Equivalently, 
\begin{align}Z\left(\frac{k}{N},\tau\right)=\sum_{Q_0=0}^{N-1} e^{-2\pi i Q_0 \frac{k}{N}} Z_{Q_0}{(\tau)}.\label{pfuncch2}\end{align}

A simple candidate for a $\mathbb{Z}_N$ version of the WGC in the parity-symmetric case would be the statement that $Z_{Q_0}(\tau)$ is bigger than $\exp\left(-\beta \frac{Q_0^2}{N}\right)$, in the low temperature $\beta\rightarrow\infty$ limit. In this limit,  $Z_{Q_0}\approx \exp(-\beta \Delta_{Q_0})$, where $\Delta_{Q_0}$ is the dimension of the lowest dimension operator in the sector of charge $Q_0\mod N$. Therefore, $Z_{Q_0}>\exp\left(-\beta \frac{Q_0^2}{N}\right)$ would mean that we have an operator with $h+\tilde{h}$ lower than $\frac{c}{12}+Q^2/N$. This is precisely the black hole threshold \eq{exj0}. We will see later on that this form of the conjecture is not satisfied using modular invariance alone, in Section \ref{sec:example}.

 We will now use modular invariance to relate
 \begin{align}Z\left(\frac{k}{N},\tau\right)&= \exp\left(-\pi i \frac{k}{N\tau}\right) Z\left(\frac{k}{N\tau},-\frac{1}{\tau}\right)=\text{Tr}\left( q^{L_0'+\frac{(Q-k)^2}{2N}}\bar{q}^{\tilde{L}_0}\right)=Z_{S,-k}(q),\nonumber\\ q&\equiv\exp\left(-2\pi i \frac{1}{\tau}\right).\end{align}
 In other words, as discussed in Appendix \ref{app:specflow}, modular invariance again relates the partition function with chemical potential to the partition function without chemical potential after spectral flow of the theory by $-k$ units, $Z_{S,-k}$. Using invariance of the theory under spectral flow by $N$ units, the partition function in the sector of charge $Q_0\mod N$ may be factorized as
 \begin{align}Z_{Q_0}\left(-\frac{1}{\tau}\right)&=\mathcal{Z}_{Q_0}\left(-\frac{1}{\tau}\right)\left[ \sum_{l=-\infty}^\infty \exp\left(2\pi i \tau \frac{(Q_0+lN)^2}{2N}\right)\right]\nonumber\\&=\mathcal{Z}_{Q_0}\left(-\frac{1}{\tau}\right)\sqrt{\frac{\tau}{Ni}}\vartheta\left(\frac{Q_0}{N},\frac{\tau}{N}\right).\end{align}
 $\mathcal{Z}_{Q_0}(\tau)$ is the partition function of charge exactly $Q_0$. Spectral flow allows us to recover the full partition function in the sector of charge $Q_0\mod N$ from $\mathcal{Z}_{Q_0}(\tau)$.  This allows us to rewrite the partition function of the spectrally flowed theory as 

 \begin{align}Z\left(\frac{k}{N},\tau\right)=Z_{S,-k}\left(-\frac{1}{\tau}\right)=\sum_{Q_0=0}^{N-1}\mathcal{Z}_{Q_0}\left(-\frac{1}{\tau}\right)\sqrt{\frac{\tau}{Ni}}\vartheta\left(\frac{Q_0-k}{N},\frac{\tau}{N}\right).\end{align}
 We may further expand each term into a series in $\exp(2\pi i k/N)$,
 \begin{align}\vartheta\left(\frac{Q_0-k}{N},\frac{\tau}{N}\right)=\frac{\exp\left(\frac{-\pi i Q_0^2}{N\tau}\right)}{\sqrt{- i N\tau}}\sum_{Q_1=0}^{N-1}e^{2\pi i Q_1\frac{k}{N}}\vartheta\left(\frac{Q_1}{N}+\frac{Q_0}{N\tau},-\frac{1}{N\tau}\right),\end{align}
 so that
 \begin{align}Z\left(\frac{k}{N},\tau\right)&=\nonumber\\\sum_{Q_1=0}^{N-1}&e^{-2\pi i Q_1\frac{k}{N}} \left[\frac{1}{N}\sum_{Q_0=0}^{N-1}\mathcal{Z}_{Q_0}\left(-\frac{1}{\tau}\right)\exp\left(\frac{-\pi i Q_0^2}{N\tau}\right) \vartheta\left(\frac{Q_1}{N}+\frac{Q_0}{N\tau},-\frac{1}{N\tau}\right)\right].\end{align}
 On the other hand, we also have
 \begin{align}Z\left(\frac{k}{N},\tau\right)=\sum_{Q_1=0}^{N-1} e^{-2\pi i Q_1 \frac{k}{N}} Z_{Q_1}(\tau)=\sum_{Q_0=1}^{N-1} e^{-2\pi i Q_1 \frac{k}{N}} \frac{\mathcal{Z}_{Q_1}(\tau)}{\sqrt{-i\tau N}}\vartheta\left(\frac{Q_1}{N},-\frac{1}{N\tau}\right).\end{align}
 Equating both expressions we arrive at 
 \begin{align}\mathcal{Z}_{Q_1}(\tau)&=\sum_{Q_0=0}^{N_1}M\left(-\frac{1}{\tau}\right)_{Q_1Q_0} \mathcal{Z}_{Q_0}\left(-\frac{1}{\tau}\right),\nonumber\\M(\tau)_{Q_1Q_0}&=\sqrt{\frac{i}{N\tau}}\exp\left(\frac{\pi i\tau Q_0^2}{N}\right) \frac{\vartheta\left(\frac{Q_1}{N}-\frac{Q_0}{N}\tau,\frac{\tau}{N}\right)}{\vartheta\left(\frac{Q_1}{N},\frac{\tau}{N}\right)}=\sqrt{\frac{i}{\tau}}\frac{1}{\sqrt{N}}\exp\left(2\pi i \frac{Q_0Q_1}{N}\right) .\label{modfourier}\end{align}
The matrix $M(\tau)$ implements the $S$ modular transformation on the charged partition functions. It is simply a unitary discrete Fourier transform $U$ times a nontrivial factor. 

Much like in the original derivation of Cardy's formula \cite{Cardy:1986ie,Carlip:2005zn}, we can use \eq{modfourier} to ascertain the UV behavior of the partition function in each charge sector. We expect the sector of charge $Q_0$ mod $N$ to be described in AdS by a large black hole with charge $Q_0$ mod $N$. Unlike in the previous case, this $\mathbb{Z}_N$ charge can indeed be observed from infinity via an Aharonov-Bohm experiment in which a particle of charge 1 circles the black hole. However, the charge will only be observable as long as there are no other effects which cancel the  Aharonov-Bohm phase; in the example which we will introduce in Section \ref{sec:example} this is precisely what happens.

Let us now explore the consequences of \eq{modfourier}; we will take $\tau=\frac{i\beta}{2\pi}$ in what follows, so that we have an ordinary partition function. In the $\beta\rightarrow\infty$ limit, the partition function $\mathcal{Z}_{Q_0}$ goes as $\exp(-\beta\tilde{\Delta}_{Q_0})$, where $\tilde{\Delta}_{Q_0}$ is the dimension of the operator with lowest dimension in the sector of charge $Q_0$ mod $N$, after substraction of the Sugawara contribution $\frac{Q_0^2}{2N}$ (due to spectral flow invariance, the operator of lowest dimension will actually have $U(1)$ charge $Q_0$) . Of course, $\tilde{\Delta}_{Q_0}\geq\Delta_0$, the conformal dimension of the vacuum, for unitarity. 

There are two different behaviors in the UV partition function, depending on the number of conformal dimensions $\tilde{\Delta}_{Q_0}$ equal to $\Delta_0$:\begin{itemize}
\item If $\tilde{\Delta}_{Q_0}=\Delta_{0}$ only for $Q_0=0$, then in the $\beta\rightarrow\infty$ limit the vector of partition functions looks like
\begin{align}\vec{\mathcal{Z}}=(\mathcal{Z}_0,\mathcal{Z}_1,\ldots,\mathcal{Z}_{N-1})=e^{-\beta\Delta_0}(1,0,\ldots)+\vec{c},\end{align}
where $\vec{c}$ contains exponentially small corrections. According to \eq{modfourier}, the partition functions in the $\beta\rightarrow0$ limit are simply
\begin{align}Z_{Q_0}=\frac{1}{N}\exp\left(-\frac{4\pi^2}{\beta} \Delta_0\right)+c'_{Q_0}, \label{fsds}\end{align}
where again $c'_{Q_0}$ vanish exponentially fast, i.e. they go as as $\exp(-\beta*x)$ with $x>0$.  In other words, in the high temperature limit every sector is dominated by a black hole configuration of the same action, whose action is identical to that of the usual BTZ black hole modulo a correction logarithmic in $N$. In this way, summing over every sector we recover the usual partition function for the BTZ black hole, as should be the case since all the assumptions that lead to Cardy's result still apply. This is as expected; the discrete $\mathbb{Z}_N$ hair of the black hole is washed out in the thermodynamic limit \cite{Coleman:1991ku}; Nevertheless, the phase might still be detectable in principle via an Aharonov-Bohm effect.

Notice that the $c'_{Q_0}$ in \eq{fsds} translates to an exponentially suppressed correction to the black hole effective action, which depends on the dimensions of the lowest-lying states in each sector:
\begin{align}I_{Q_0}=\ln Z_{Q_0}=-\frac{4\pi^2}{\beta}\Delta_0\left(1+\ln\left(c'_{Q_0}e^{\frac{4\pi^2}{\beta}\Delta_0}\right)\right).\end{align}
It is natural to guess that the bulk description of these nonperturbative effects corresponds to the nonperturbative contributions discussed, in the 4d context, in \cite{Coleman:1991ku}: Namely, that they correspond to instantonic euclidean particles  wrapping the $S^1$ of the euclidean black hole, so that they give a charge-dependent contribution to the action thanks to the Aharonov-Bohm phase. 

\item If $\tilde{\Delta}_{Q_0}=\Delta_{0}$ for some $Q_0\neq 0$, in the $\beta\rightarrow\infty$ limit the vector of partition functions looks like
\begin{align}\vec{\mathcal{Z}}= e^{-\beta\Delta_0}\vec{v}+\vec{c},\end{align}
where $\vec{v}$ has a 1 for every $Q_0$ with $\Delta_{Q_0}=\Delta_{0}$ and $\vec{c}$ again contains corrections exponentially smaller than the first term. Unlike in the previous case, the partition functions of every black hole are not identical, but rather we have
\begin{align}Z_{Q_0}=g_{Q_0}\exp\left(-\frac{4\pi^2}{\beta} \Delta_0\right)+c'_{Q_0}, \label{fsds2}\end{align}
for some constants $g_{Q_0}$. Thus, the effective actions of the charged black holes are different from each other, by a logarithmic correction. The sum
\begin{align}Z=\sum_{Q_0}Z_{Q_0}=\exp\left(-\frac{4\pi^2}{\beta}\Delta_0\right)\end{align}
still reproduces Cardy's result, due to the properties of the discrete Fourier transform.

In the extreme case where $\vec{v}=(1,1,\ldots,1)$, the $Q_0=0$ black hole has an entropy equal to the usual BTZ black hole, while the high temperature behavior of the partition function of every other charge sector is not dominated by the operator of lowest dimension, but rather by the one of second-lowest one. This certainly does not coincide with the results from charged black holes with a Chern-Simons term, so that this case is not likely to be dual to a weakly coupled $U(1)$+gravity in the bulk.
\end{itemize}
We therefore see that modular invariance relates the presence of extra particles saturating the unitarity bound in the $Q_0\neq0$ sector with relatively large corrections to the charged black hole free energy (i.e. corrections not exponentially suppressed). Black hole free energy computed from the weakly coupled bulk dual strongly suggests that we are in the first case described above; as a result, no particles saturating the bound are likely to exist.

This does not preclude the existence of charged operators with conformal dimension below the black hole threshold \eq{exj0}; as discussed above such operators would satisfy the WGC. In what follows we will provide a simple example in which this is not the case, as well as discuss generic bounds to the dimension of these charged operators via modular bootstrap of \eq{modfourier}. This will allow us to connect with the results of \cite{Benjamin:2016fhe} in an explicit way.

\subsection{An example based on alignment}\label{sec:example}
The considerations of the previous section suggest that, unlike in Section \ref{sec:specflow}, modular invariance is not directly concerned with the existence of $\mathbb{Z}_N$ charged states of low conformal dimension. We will now provide an explicit example thus showing that this is indeed the case: The state of lowest $\mathbb{Z}_N$ charge can have an arbitrarily high conformal dimension. 

Consider a modular invariant theory consisting of just two left-moving and two right-moving $U(1)$'s at level 1. The charge lattice is the even self dual $\Lambda_{2,2}$, spanned by the vectors $\{(1,0,-1,0),(\frac12,0,\frac12,0), (0,1,0,-1),(0,\frac12,0,\frac12)\}$.
With this choice, the $U(1)^2$ current algebra is dual to two bosons compactified at the self-dual radius; the associated factor in the partition function is modular invariant by itself. The theory has central charge $c=2$; we can take the tensor product with any other theory of our choice to achieve the large central charge regime necessary to have a weakly coupled gravitational dual. 

A generic vector in the charge lattice will be labeled as $(a,b,c,d)$. The left-moving Sugawara $L_0$ is 
\begin{align} L_0^S=\frac12(a^2+b^2).\end{align}
There are current algebras of any level  we desire embedded here. For instance, consider the lattice vector $\vec{v}=(N,1,0,0)$; in what follows we will omit the last two entries as we will be working purely within the left-moving part of the algebra. A vector $(a,b)$ in the charge lattice may be decomposed into a component along $\vec{v}$ and another $\vec{v}_\perp$ orthogonal to it. The longitudinal component is the vector
\begin{align}\frac{Q}{N^2+1}(N,1),\quad Q\equiv aN+b.\end{align}
so that $L_0^S$ takes the form
\begin{align}L_0^S=\frac{Q^2}{2(N^2+1)}+\frac{\vert \vec{v}_\perp\vert ^2}{2}.\end{align}
Thus, we have rewritten the system as a $U(1)$ current algebra at level $N^2+1$, plus an extra contribution. Actually, $Q$ is quantized in half-integers, and thus in the conventions used in the rest of the paper (where charges are integer quantized) the level would actually be $4(N^2+1)$. We want to know what is the lowest possible value of the extra term, $\frac{\vert \vec{v}_\perp\vert ^2}{2}$, in each sector of charge $Q$. Since
\begin{align}\vert \vec{v}_\perp\vert=\frac{1}{\sqrt{N^2+1}}\left\vert QN -(N^2+1)a\right\vert,\end{align}
the lowest value is attained when $a$ is such that $\vert \frac{QN}{N^2+1} -a\vert\leq 1/4$, or in other words
\begin{align}\frac{\vert \vec{v}_\perp\vert ^2_{\text{min}}}{2}=\frac{N^2+1}{8}\text{dist}^2\left( \frac{2QN}{N^2+1}\right),\label{ffdsafa}\end{align}
where $\text{dist}^2(x)$ is the square of the distance of the real number $x$ to the nearest integer. 

For $Q=1$, this reduces to $N^2/(2(N^2+1))$, which is always smaller than $1/2$. Thus, the correction (over the Sugawara term) in this sector (charge $Q_0=1$ mod $N$) is always relatively small. However, the maximum of the above equation for large $N$  grows as $N^2/32$, since for large $N$ there is enough density so that for some $Q\approx \frac{N}{2}(a+1/2)$, $2NQ/(N^2+1)$ is close to a half-integer, so that the $\text{dist}^2(x)$ term in \eq{ffdsafa} is just $1/4$.

Thus, in this particular example, we see that we can get a level $N$ current algebra in which the dimension of the lowest state in some charged sector has a gap of at least $N^2/32$. We can do this at fixed central charge, which implies that by a perverse choice of $U(1)$ we can have a current algebra for which the state of lowest charge in some sector is as massive as we want. This is in direct violation of the Lattice WGC which, as discussed in Section \ref{blackholesols}, is the  version of the conjecture expected to apply in AdS backgrounds. This specific example thus proves that modular invariance alone is not enough to prove this version of the Lattice WGC for theories with a compact current algebra at level $N$.

We can also see invariance under spectral flow explicitly. The non-Sugawara contribution to $L_0$ is given by
\begin{align}\frac{\vert \vec{v}_\perp\vert ^2}{2}=\frac{1}{2(N^2+1)}\left(QN -(N^2+1)a\right)^2.\label{termvpar}\end{align}
Equation \eq{termvpar} is invariant under $Q\rightarrow 2(N^2+1)$, $a\rightarrow a+2N$. This corresponds to shifting $(a,b,c,d)$ by $(2N,2(N^2-N+1),0,0)$, which is a vector in the charge lattice, and thus this transformation is allowed. Every state in the sector of charge $Q$ is thus mapped to another with the same $\frac{\vert \vec{v}_\perp\vert ^2}{2}$ but different $Q$, as demanded by spectral flow invariance.

The above example seems to rule out any version of the WGC for $\mathbb{Z}_N$ theories, having to settle for the spectral flow version discussed in Section \ref{sec:specflow}. However, this particular mechanism to engineer a level $N$ current algebra has the drawback that the underlying charge lattice is always an integer lattice. When performing an Aharonov-Bohm phase experiment in which a particle with charge vector $\vec{Q}_p$ circles a black hole with charge vector $\vec{Q}_{\text{BH}}$, the resulting phase is proportional to $\vec{Q}_p\cdot \vec{Q}_{\text{BH}}$ which is an integer and thus trivial. The Aharonov-Bohm phase resulting from the charge under $U(1)_{\text{v}}$ is cancelled by the contribution coming from the orthogonal $U(1)_{\text{v}_\perp}$. Hence there is no way to distinguish black holes with different $\mathbb{Z}_{4(N^2+1)}$ charge, and as a result, there is no WGC argument to begin with. 

In light of these considerations, it is entirely possible that some version of the WGC actually holds for $\mathbb{Z}_N$ theories in which the black hole $\mathbb{Z}_N$ charge is actually measurable at infinity via an Aharonov-Bohm experiment. Nevertheless, unlike for the spectral flow WGC in Section \ref{sec:specflow}, modular invariance is not enough. 
\subsection{Modular bootstrap approach}\label{sec:mboost} 
Equation \eq{modfourier} is well suited for the modular bootstrap approach pioneered in this context in \cite{Hellerman:2009bu}. The recent work \cite{Benjamin:2016fhe} used modular bootstrap techniques to place constraints on the dimension of the charged operator of lowest weight in a theory with an abelian current algebra. The results in this reference are quite general and work even for noncompact gauge groups. For compact groups, we have seen that spectral flow provides us with stronger bounds, as long as the current algebra level is $N=1$. For higher $N$, there are no such constraints, as we have discussed in Sections \ref{modznconsts} and \ref{sec:example}. However, the bound of \cite{Benjamin:2016fhe} must still apply, and it would be interesting to figure out to which charge mod $N$ do they apply. It is clear from the example in Section \ref{sec:example} that modular invariance alone is not enough to prove the WGC; in this Section however we will explore the constraint placed by modular invariance on the spectrum.

Following \cite{Hellerman:2009bu}, let us look at modular invariance near $\tau=i$, by parametrizing $\tau=i\exp(s)$. Then, the modular invariance constraint \eq{modfourier} takes the form 
\begin{align}\vec{\mathcal{Z}}(-s)=\exp(-s/2) U \vec{\mathcal{Z}}(s).\label{modb}\end{align}
This is especially well suited for modular bootstrap purposes. Using \eq{modb} twice, we see that $ \vec{\mathcal{Z}}(s)=U \vec{\mathcal{Z}}(s)$, so that $ \vec{\mathcal{Z}}(s)$ decomposes in two pieces, $ \vec{\mathcal{Z}}(s)=\vec{\mathcal{Z}}_+(s)+\vec{\mathcal{Z}}_-(s)$ with eigenvalues $\pm1$ under $U$ (the discrete Fourier transform operator has in general four distinct eigenvalues $\pm1,\pm i$; the latter are absent in this context). Since $U$ is unitary, $\vec{\mathcal{Z}}_\pm(s)$ are orthogonal and \eq{modb} becomes
\begin{align}\vec{\mathcal{Z}}_\pm(-s)=\pm \exp(-s/2) \vec{\mathcal{Z}}_\pm(s).\label{modb2}\end{align}

Let us take the $n$-th derivative of \eq{modb2} and evaluate at $s=0$ (we will omit the argument  for convenience):
\begin{align}(-1)^n\vec{\mathcal{Z}}_\pm^{(n)}=\pm\sum_{k=0}^n\binom{n}{k}\left(-\frac12\right)^k \vec{\mathcal{Z}}_\pm^{(n-k)}.\end{align}
Derivatives with respect to $s$ are related to those with respect to $\beta$ as
\begin{align}\mathcal{Z}^{(n)}_{Q_0}=\sum_{k=1}^n \genfrac\{\}{0pt}{}{n}{k}(2\pi)^k\frac{d^k\mathcal{Z}}{d\beta^k}=\sum_{k=1}^n\genfrac\{\}{0pt}{}{n}{k}(-2\pi)^k\langle \Delta^k\rangle_{Q_0} \mathcal{Z}_{Q_0},\end{align}
where $\genfrac\{\}{0pt}{}{n}{k}$ are Stirling numbers of the second kind.

The constraints \eq{modb2} can be attacked in principle via linear programming techniques, supplemented with the constraints $\langle \Delta_{Q_0}^k\rangle\geq \Delta_0^k$ which ensure that the conformal dimension of every operator is above the vacuum $\Delta_0$.  We will present here the results for $n=1$ only. The variables are $\vec{\mathcal{Z}}_\pm,\vec{\mathcal{Z}}_\pm^{(1)}$, and the constraints are
\begin{align}\vec{\mathcal{Z}}_-=0,\quad\vec{\mathcal{Z}}_+^{(1)}=\frac14\vec{\mathcal{Z}}_+.\end{align}
This means that 
\begin{align}\vec{\mathcal{Z}}^{(1)}=\vec{\mathcal{Z}}_-^{(1)}+\frac14\vec{\mathcal{Z}}_+,\quad \vec{\mathcal{Z}}=\vec{\mathcal{Z}}_+,\end{align}
and as a result
\begin{align}-2\pi \langle E_{Q_{0}}\rangle=\frac14+\frac{\mathcal{Z}_{-,Q_0}^{(1)}}{\mathcal{Z}_{+,Q_0}}.\end{align}
We are interested in an upper bound for $\langle E_{Q_{0}}\rangle$, so all we have to do is minimize the above equation where $\vec{\mathcal{Z}}_{-}^{(1)}$ is a generic vector of the negative eigenspace of the Discrete Fourier Transform, and $\vec{\mathcal{Z}}_{+}$ is a generic vector of the positive eigenspace, with all its components positive. This is a very simple linear programming problem, if we take $\mathcal{Z}_{+,Q_0}=1$, which we can do without loss of generality. We have to supplement the above with the constraint that the energy of no sector can be lower than that of the vacuum,
\begin{align} \left(-2\pi \Delta_0-\frac14\right)\mathcal{Z}_{+,Q_0}\geq \mathcal{Z}_{-,Q_0}^{(1)}.\end{align}
For $N=2$, these constraints can be easily solved analytically to yield a bound
\begin{align}\langle \Delta_1\rangle \leq  \frac{\left(3+2 \sqrt{2}\right) \pi \frac{c+\tilde{c}}{2}-3 \left(2+\sqrt{2}\right)}{12 \pi },\end{align}
Thus, there must be some operator with charge 1 mod 2 below the right hand side of the above equation. For large central charge, this bound is above the  threshold \eq{wgc-mod-bootstrap}. The bound is also less stringent than \eq{wgc-mod-bootstrap}.

For $N=3$, one can also solve analytically the problem, to obtain 
\begin{align}\langle \Delta_1\rangle =\langle \Delta_2\rangle\leq\frac{2 \left(2+\sqrt{3}\right) \pi  \frac{c+\tilde{c}}{2}-3 \left(3+\sqrt{3}\right)}{24 \pi }\end{align}
For $N=4$ and beyond, however, the linear programming problem is unbounded, save for the (uninteresting) vacuum sector $Q_0=0$. Thus, it is not possible to get nontrivial constraints in this case, at least with the simple approach taken here.

\section{Conclusions}\label{conclus}

The WGC is a very useful and seemingly general feature of stringy models, which is also supported by some general quantum gravity arguments, though these present several loopholes and caveats. We have explored the WGC in three-dimensional AdS space, where most of these arguments do not directly apply, and found that nonetheless a version of the (Lattice) WGC conjecture seems to hold in the context of the AdS/CFT correspondence. More specifically, we have shown that in any theory with a compact $U(1)$ and a modular invariant holographic dual there are light charged states below the black hole threshold, which, in a certain sense, plays the role of the 3d extremality bound. 

Our considerations make manifest that, at least in the three-dimensional case, the WGC is independent of any remnant-based arguments. It is still true that the light states predicted by the WGC allow for a Hawking-Page phase transition for small black holes in charged sectors. 

The electric charge of a large black hole in AdS is represented by a flat holonomy, which is not measurable semiclassically. Neither Aharonov-Bohm type of experiments nor the semiclassical entropy formula seem sensitive to it (aside from a ${\mathbb Z}_N$ subgroup). This peculiarity of three-dimensional black holes renders gauge charges in 3d very similar to global charges, at least in the semiclassical limit. We know however that electric charge is exactly conserved, as a consequence of the extended chiral algebra of the CFT. The usual arguments against global symmetries do not apply, both because large AdS black holes do not evaporate and because we expect significant quantum corrections for any black hole smaller than the AdS radius.

Consistent coupling of a $U(1)$ in three dimensions requires a Chern Simons coupling, which is dual to a current algebra at level $N$ in the CFT. In this case, modular invariance is enough to show that the WGC holds for the quotient $U(1)/\mathbb{Z}_N$, where $\mathbb{Z}_N$ is generated by $\exp(2\pi i/N)$ in the original $U(1)$.  In other words, we only have WGC states in sectors of charge an integer multiple of $N$. We have explored the constraints of modular invariance on these $\mathbb{Z}_N$ charged sectors, showing that $\mathbb{Z}_N$ charged states saturating the unitarity bound \eq{unit-bound} mean order 1 differences in the  entropy of large black holes with different $\mathbb{Z}_N$ charge; if these states are absent then the different $\mathbb{Z}_N$ black hole entropies are equal to each other up to exponentially small corrections. We have also shown via an explicit counterexample that modular invariance is not enough to have a WGC statement in these sectors as well. Nevertheless, modular invariance does produce some nontrivial constraints for low $N$, which we have studied via a modular bootstrap approach.

To sum up, modular invariance of the holographic dual together with compactness of the gauge group seems to lead to a WGC in three dimensions. In three dimensions, just having a CFT dual is not enough for our version WGC to hold: we need to impose both modular invariance and compactness of the gauge group. This fits in nicely with the results of \cite{Nakayama:2015hga}, which studied the WGC in four-dimensional AdS space, and found that some CFT's apparently violated the conjecture. Perhaps some extra constraint, akin to modular invariance, has to be imposed on the CFT. 

Of course, the drastic differences in the behavior of both gauge fields and gravity between three and more dimensions, which played an essential role in our reasoning, may invalidate any comparison between the three-dimensional and higher dimensional cases. Whether or not this is the case, we find interesting that the proof of existence of light charged states in $AdS_3$ seems to be so independent of any considerations of remnants, a species problem, or any of the original WGC rationale.

\paragraph{Acknowledgements}
We thank  J. L. F. Barb\'{o}n,  Jon Brown, William Cottrell, Javier Mart\'{i}n, Matthew Reece, \'{A}ngel Uranga, and Gianluca Zoccarato for useful discussions and comments. This work is partially supported by the grants FPA2015-65480-P from the MINECO, the ERC Advanced Grant SPLE under contract ERC-2012-ADG-20120216-320421 and the grant SEV-2012-0249 of the ``Centro de Excelencia Severo Ochoa" Programme. M.M. is supported by a ``La Caixa'' Ph.D scholarship.  G.S. and P.S. are supported in part by the DOE grant DE-FG-02-95ER40896, the Kellett Award of the University of Wisconsin, and the HKRGC grants HUKST4/CRF/13G, 604231 and 16304414.

\appendix

\section{Duality between Chern-Simons and current algebra}\label{dualcurrent}
In this appendix we review the relationship between a Chern-Simons bulk term and a current algebra at level $N$. The following discussion is taken from \cite{Kraus:2006wn,Jensen:2010em}; we include it mostly to fix normalization. The discussion is in Euclidean signature; a similar argument runs in the Lorentzian case, taking lightcone coordinates instead of complex ones.

We start with a Chern-Simons term \eq{csterm}
\begin{align}S_{CS}=i\frac{N e^2}{4\pi}\int F\wedge A. \label{cs1}\end{align}
 
 The general AdS/CFT dictionary instructs us to evaluate variations of the on-shell bulk action to obtain the dual CFT source \cite{Witten:1998qj}. We want to obtain the current generating the global part of the $U(1)$ bulk gauge symmetry, so we should vary the on-shell action with respect to variations of the gauge field which do not vanish at infinity:
\begin{align}\delta I=\frac{ie}{2\pi }\int_{\partial AdS}d^2x j^\alpha \delta A_\alpha.\label{c2}\end{align}
We need to know what are the allowed variations $\delta A_\alpha$. In the bulk, we know that the holonomy of $A$ on the circle at infinity measures the electric charge, and as such $A_\phi$ is constrained. Hence, only the time component of $A$ is allowed to vary freely in a solution of the equations of motion. Introducing a holomorphic coordinate $w\equiv \phi + i t/l$, the variational principle holds that either $A_w$ or $A_{\bar{w}}$ must be kept fixed, but not both, when looking for solutions to the classical equations of motion. We will take $\delta A_{\bar{w}}=0$ and justify later why this is the only correct choice. That means that when varying the action there should be no boundary dependence on $\delta A_w$. In other words, \eq{c2} takes the form
\begin{align}\delta I=\frac{ie}{2\pi}\int_{\partial AdS}\sqrt{g}d^2w j^{\bar{w}} \delta A_{\bar{w}}.\label{c3}\end{align}
However, the variation of the Chern-Simons term \eq{csterm} is not of the form \eq{c3}. Explicitly, we have a boundary term
\begin{align}\delta S_{CS}^{\text{boundary}}= i\frac{Ne^2}{4\pi}\int_{\partial AdS}\delta A\wedge A= i\frac{Ne^2}{4\pi}\int_{\partial AdS}(\delta A_w A_{\bar{w}}-\delta A_{\bar{w}} A_w) dw\wedge d\bar{w}.\end{align}
This can be fixed if we add a boundary term\footnote{We are taking $*d\phi=\frac{dt}{l}$ in the euclidean theory.}
\begin{align}I_{\text{boundary}}=\frac{Ne^2}{8\pi}\int_{\partial{AdS}} A\wedge *A=-i\frac{Ne^2}{4\pi}\int_{\partial{AdS}} (A_{w}A_{\bar{w}})dw\wedge d\bar{w} \label{bterm}\end{align}
to the action. Then
\begin{align}\delta I_{\text{boundary}}=-i\frac{Ne^2}{4\pi}\int_{\partial AdS}(\delta A_w A_{\bar{w}}+\delta A_{\bar{w}} A_w) dw\wedge d\bar{w}\end{align}
Since $dw\wedge d\bar{w}= -i d^2w$, we have
\begin{align}j_w=\frac12j^{\bar{w}}= i N e A_w.\end{align}
We thus get a holomorphic operator $j_w\equiv j(w)$, of spin one and conformal dimension one. This will generate in the CFT the global part of the $U(1)$ gauge group. 

We can compute the commutator of the currents using \eq{c2} and the standard AdS-CFT prescription. We have
\begin{align} \langle \delta j\rangle=i N e \langle\delta A_w\rangle=\frac{ie}{2\pi}\int \sqrt{g}d^2w' \langle j(w') j(w)\rangle \delta A_{\bar{w}}\label{c5}\end{align}
We can evaluate $\delta A_w$ from the bulk equation of motion which tells us that, at infinity,
\begin{align}\delta F=0\quad\Rightarrow\quad \partial_{\bar{w}} \delta A_w= \partial_{w} \delta A_{\bar{w}}\end{align}
so that, taking derivatives from \eq{c5}
\begin{align}N\partial_{\bar{w}} \delta A_w=N\partial_{w} \delta A_{\bar{w}}=\frac{1}{4\pi}\int d^2 w' \langle j(w')j(w)\rangle \delta A_{\bar{w}}.\label{c6}\end{align}
This, together with the identity $\partial_{\bar{w}}(1/(2\pi w^2))=-\partial_w(\delta^{(2)}(w,\bar{w}))$ and the fact that the above expectation values hold with arbitrary additional operator insertions specifies uniquely the OPE
\begin{align}j(w)j(0)=\frac{N}{w^2 }+\text{holomorphic terms}.\label{calbebra}\end{align}
This is the OPE of a current algebra at level $N$.  The modes
\begin{align}j_n=\int \frac{dw}{2\pi i}e^{-i n w} j(w)\end{align}
satisfy the commutation relations $[j_m, j_n]=N\delta_{m+n}$. Notice that the zero mode
\begin{align}j_0=\frac{eN}{2\pi}\int_{S^1} A\end{align}
is, by virtue of \eq{qishol}, precisely the electric charge of the corresponding bulk configuration, in multiples of $e$.

\section{Constraints on the partition function from modular invariance}\label{app:specflow}
Here we sketch a proof of the statement in Section \ref{sec:specflow} that modular invariance of a CFT with a $U(1)$ current algebra implies invariance under spectral flow. While this claim has been made before \cite{ArkaniHamed:2006dz}, we are not aware of a detailed presentation of the argument in the literature. We also provide a proof of the modified Cardy formula \eq{cformain}, along the lines of the original treatment.

\subsection{Spectral flow from modular invariance}
In theories with a holomorphic current $j(w)$, the partition function on the torus should be invariant not only under modular transformations, but also under $U(1)$ transformations.  In two dimensions, the conformal algebra enlarges the symmetry group of the theory so that local transformations are also symmetries. A transformation with parameter $e^{i\lambda(w)}$ is generated by \eq{c2}, with $\delta A_{\bar{w}}=\partial_{\bar{w}}\lambda$. Integration by parts yields the Ward identity \cite{Polchinskiv2}
\begin{align}\delta \mathcal{A}(w_0)= i\lambda(w_0) [Q,\mathcal{A}(w_0)]=iQ_{\mathcal{A}}\lambda(w_0)\mathcal{A}(w_0)\label{ss2a}\end{align}
where we have assumed the operator to have definite charge $Q_{\mathcal{A}}$. The finite form of \eq{ss2a} gives the transformation of a generic local operator,
\begin{align}\mathcal{A}(w_0)\rightarrow e^{i Q\lambda(w_0)}\mathcal{A}(w_0).\label{wid}\end{align}
We will be interested in the translation of the local operator \eq{wid} by some number $z$. The generator $Q$ commutes with momentum $P=L_0-\tilde{L_0}$ and the Hamiltonian $L_0+\tilde{L}_0$ due to the extended algebra \eq{eca}, so that an infinitesimal translation along direction $2\pi z$ acts on on \eq{wid} as
\begin{align}\left.\frac{d(U_{zt} e^{i\lambda(w_0) Q}\mathcal{A}(w_0)U_{tz}^{-1})}{dt}\right\vert_{a=0}= \frac{i}{2} Q (z\partial_w\lambda(w_0)+\bar{z}\partial_{\bar{w}}\lambda(w_0))\mathcal{A}(w_0) + \partial \mathcal{A}(w_0).\end{align}
Integration of this differential equation gives us the translated \eq{wid} by $z$,
\begin{align} U_{zt} e^{i\lambda(w_0) Q}\mathcal{A}(w_0)U_{tz}^{-1}=e^{i\left(\lambda(w_0)+\frac{i}{2} z\int_{w_0}^{w_0+z} (\partial_{w}\lambda(w) dw+\partial_{\bar{w}}\lambda(w) d\bar{w})\right) Q}\mathcal{A}(w_0).\label{gwid}\end{align}
Equations \eq{wid} and \eq{gwid} are the same in contractible spaces, but in general they do not coincide. Consider the theory on a torus. There, there are  translations along non-contractible cycles which return the operator to the same point. The map $\lambda(w)$ may have winding along this cycle. The global structure of the gauge group is related to the allowed windings of the map $\lambda(w)$ on the spatial cycle of the torus. If the gauge group is $\mathbb{R}$, no winding is allowed, but if it is compact, the map $\lambda(w)=\pi(w+\bar{w})$, which winds along the $\phi$ cycle, is allowed. From comparison of \eq{gwid} and \eq{wid}, we find that
\begin{align}\exp(2\pi i Q) \mathcal{A}(w_0)=\mathcal{A}(w_0)e^{i\phi_{\mathcal{A}(w_0)}} \label{chargeq}\end{align}
where $\phi_{\mathcal{A}(w_0)}$ is some phase. Consistency with e.g. the OPE of the currents then implies that this phase is trivial. Equation \eq{chargeq} then means that $Q$ is quantized. One may run the argument backwards to show that quantization of the electric charge means that the theory is invariant under gauge transformations with integer winding along the $\phi$ cycle, and thus we  recover the familiar result that charge quantization is equivalent to a compact gauge group.

Taking $\mathcal{A}(w_0)=T_{ww}(w_0)$ and integrating over $\phi$ one obtains $e^{2\pi i Q} L_0=L_0$. Exponentiation then yields $e^{2\pi i Q}q^{L_0}=q^{L_0}$.  If we define, following \cite{Benjamin:2016fhe}
\begin{align}Z(z,\tau)=\text{Tr}\left(q^{L_0-\frac{c}{24}}\bar{q}^{\tilde{L}_0-\frac{\tilde{c}}{24}}e^{2\pi i z Q}\right)\end{align}
invariance under $U(1)$ transformations with winding amounts to $Z(z,\tau)=Z(z+1,\tau)$. As above, a necessary and sufficient condition for this is that every state in the theory has quantized charges. 

So far we have said nothing about $\lambda(w)$ having winding on the $\tau$ cycle. Invariance under this transformation is not directly related to the compactness of the gauge group. However, it is a symmetry of the partition function on a torus, as we will now see. Upon a generic modular transformation one has
\begin{align}\tau\rightarrow\frac{a\tau+b}{c\tau+d},\quad z\rightarrow\frac{z}{c\tau+d}.\label{modtr}\end{align}
Modular transformations relate a configuration with winding along the $\sigma$ cycle to another one with winding line along the $\tau$ cycle. 
Furthermore, the partition function $Z(\tau,z)$ transforms under \eq{modtr} according to the universal rule \cite{Kraus:2006wn,Benjamin:2016fhe}
\begin{align}Z(z',\tau')= \exp\left(i\pi N\frac{cz^2}{c\tau+d}\right) Z(z,\tau).\label{modprop}\end{align}

Our expression differs from that of \cite{Benjamin:2016fhe} because we are considering a purely holomorphic current; conjugate terms should be added in case of a current with holomorphic and antiholomorphic components. Consider now the series of transformations consisting of an $S$-transformation, a $z\rightarrow z+1$ shift, and another $S$-transformation. This takes $z$ to $z+\tau$ while leaving $\tau$ invariant. Applying this to $Z(\tau,0)$, one obtains
\begin{align}Z(\tau,\tau)=\exp(-i\pi N\tau)Z(0,\tau).\end{align}
This encodes the transformation properties of the partition function under symmetry transformations with winding along the time direction. 
We may rewrite this as
\begin{align}Z(0,\tau)=\text{Tr}\left(q^{L_0-\frac{c}{24}}\bar{q}^{\tilde{L}_0-\frac{\tilde{c}}{24}}\right)=\text{Tr}\left(q^{L_0-\frac{c}{24}+Q+\frac{N}{2}}\bar{q}^{\tilde{L}_0-\frac{\tilde{c}}{24}}\right)=Z(\tau,\tau).\end{align}
The only way this can hold for every possible value of $\tau$ is if 
\begin{align}L_0=\frac{Q^2}{2N}+L_0'\quad\text{and}\quad L_0+Q+\frac{N}{2}=\frac{(Q+N)^2}{2N}+L_0'\end{align}
have the same spectrum, and are thus related by a unitary transformation $U_N$ (which commutes with $\bar{L}_0$\footnote{Since the above transformation shifts $h$ by $Q+\frac{N}{2}$ while leaving $\tilde{h}$ invariant.} ) . Applying this transformation twice, we get
\begin{align}\frac{(Q+2N)^2}{2N}+L_0'= U_N\left( \frac{(Q+N)^2}{2N}+L'_0\right)\end{align}
which implies $U_N Q U_{N}^{-1}=Q+N$. 

Thus, the operator $U_N$ corresponds precisely to the spectral flow automorphism \eq{specflow} by $N$ units. Hence modular invariance requires the spectrum to be invariant under a change of $Q$ by $N$ units and the left-moving weight $h$ by $Q+\frac{N}{2}$, while leaving $\tilde{h}$ invariant. 
 
\subsection{Cardy's formula for the charge \texorpdfstring{$Q$}{Q} sector}\label{app:cardy}
We will now use the modular invariance constraints from the previous section to obtain a formula for the high temperature limit of the partition function in the charge $Q$ sector. We will discuss the parity-symmetric case, so that there are both left and right-moving currents at the same level. From \eq{modprop}, and taking $\tau=\frac{i\beta}{2\pi}$, we have, using invariance under spectral flow,
\begin{align}Z\left(z,\frac{i\beta}{2\pi}\right)=Z_{S,-zN}\left(0,\frac{2\pi}{\beta} i\right).\end{align}
If we assume the existence of a gap between the vacuum and the first massive excitation, in the spirit of Cardy's result, then in the $\beta\rightarrow0$ limit the right hand side of the above equation is dominated by the operator of lowest dimension, after spectral flow. In general, this will be different for different values of $z,\tilde{z}$, depending on the spectrum. However, for small enough $z,\tilde{z}$, the dominant state will still be the vacuum, and thus
\begin{align} Z\left(z,\frac{i\beta}{2\pi}\right)\approx \exp\left( -\frac{4\pi^2}{\beta}\left(\Delta_0+ N\frac{z^2+\tilde{z}^2}{2}\right)\right)\end{align}
Similarly, due to invariance under spectral flow by $N$ units, for $z,\tilde{z}$ close enough to 1
\begin{align} Z\left(z,\frac{i\beta}{2\pi}\right)\approx \exp\left( -\frac{4\pi^2}{\beta}\left(\Delta_0+ N\frac{(z-1)^2+(\tilde{z}-1)^2}{2}\right)\right).\end{align}
For $z=k/N$, we could have another dominant contribution if there was a particle of $\mathbb{Z}_N$ charge $k$ saturating the inequality \eq{unit-bound}. However, as discussed in Section \ref{modznconsts}, saturation of the inequality is incompatible with the semiclassical limit; therefore, although $Z\left(z,\frac{i\beta}{2\pi}\right)$ may have additional peaks at $z=k/N$, the operator of lowest dimension in the spectrally flowed theory has $-\Delta_0+\epsilon$, with $\epsilon>0$; as a result, in the $\beta\rightarrow0$ limit, these peaks are much lower than the ones at $z=0,1$. The same analysis is valid for $\tilde{z}$, and we arrive at the conclusion that in the $\beta\rightarrow0$ limit we can approximate
\begin{align} Z\left(z,\frac{i\beta}{2\pi}\right)&\approx \exp\left( -\frac{4\pi^2}{\beta}\Delta_0\right)\nonumber\\&\cdot \left[\exp\left( -\frac{4\pi^2}{\beta} N\frac{z^2+\tilde{z}^2}{2}\right)+\exp\left( -\frac{4\pi^2}{\beta} N\frac{(z-1)^2+(\tilde{z}-1)^2}{2}\right)\right].\end{align}
Now that we have an approximate expression for the grand canonical partition function with imaginary chemical potential, it is trivial to get the partition function in the charge $Q$ sector via Fourier transform:
\begin{align}Z_{Q}\left(z,\frac{i\beta}{2\pi}\right)=\int_0^1 d\tilde{z}\int_0^1 dz\  Z\left(z,\frac{i\beta}{2\pi}\right)e^{2\pi i Qz}=\frac{\beta}{2\pi N}\exp\left( -\frac{4\pi^2}{\beta}\Delta_0-\frac{Q^2}{N}\beta\right),\end{align}
which gives \eq{cformain} up to a prefactor describing logarithmic corrections to the free energy.

\bibliographystyle{JHEP}
\bibliography{wgc3drefs}

\providecommand{\href}[2]{#2}\begingroup\raggedright\begin{thebibliography}{10}

\bibitem{ArkaniHamed:2006dz}
N.~Arkani-Hamed, L.~Motl, A.~Nicolis, and C.~Vafa, {\it {The String landscape,
  black holes and gravity as the weakest force}},  {\em JHEP} {\bf 06} (2007)
  060, [\href{http://arxiv.org/abs/hep-th/0601001}{{\tt hep-th/0601001}}].

\bibitem{Rudelius:2014wla}
T.~Rudelius, {\it {On the Possibility of Large Axion Moduli Spaces}},  {\em
  JCAP} {\bf 1504} (2015), no.~04 049,
  [\href{http://arxiv.org/abs/1409.5793}{{\tt arXiv:1409.5793}}].

\bibitem{delaFuente:2014aca}
A.~de~la Fuente, P.~Saraswat, and R.~Sundrum, {\it {Natural Inflation and
  Quantum Gravity}},  {\em Phys. Rev. Lett.} {\bf 114} (2015), no.~15 151303,
  [\href{http://arxiv.org/abs/1412.3457}{{\tt arXiv:1412.3457}}].

\bibitem{Rudelius:2015xta}
T.~Rudelius, {\it {Constraints on Axion Inflation from the Weak Gravity
  Conjecture}},  {\em JCAP} {\bf 1509} (2015), no.~09 020,
  [\href{http://arxiv.org/abs/1503.00795}{{\tt arXiv:1503.00795}}].

\bibitem{Montero:2015ofa}
M.~Montero, A.~M. Uranga, and I.~Valenzuela, {\it {Transplanckian axions!?}},
  {\em JHEP} {\bf 08} (2015) 032, [\href{http://arxiv.org/abs/1503.03886}{{\tt
  arXiv:1503.03886}}].

\bibitem{Brown:2015iha}
J.~Brown, W.~Cottrell, G.~Shiu, and P.~Soler, {\it {Fencing in the Swampland:
  Quantum Gravity Constraints on Large Field Inflation}},  {\em JHEP} {\bf 10}
  (2015) 023, [\href{http://arxiv.org/abs/1503.04783}{{\tt arXiv:1503.04783}}].

\bibitem{Bachlechner:2015qja}
T.~C. Bachlechner, C.~Long, and L.~McAllister, {\it {Planckian Axions and the
  Weak Gravity Conjecture}},  \href{http://arxiv.org/abs/1503.07853}{{\tt
  arXiv:1503.07853}}.

\bibitem{Hebecker:2015rya}
A.~Hebecker, P.~Mangat, F.~Rompineve, and L.~T. Witkowski, {\it {Winding out of
  the Swamp: Evading the Weak Gravity Conjecture with F-term Winding
  Inflation?}},  {\em Phys. Lett.} {\bf B748} (2015) 455--462,
  [\href{http://arxiv.org/abs/1503.07912}{{\tt arXiv:1503.07912}}].

\bibitem{Brown:2015lia}
J.~Brown, W.~Cottrell, G.~Shiu, and P.~Soler, {\it {On Axionic Field Ranges,
  Loopholes and the Weak Gravity Conjecture}},  {\em JHEP} {\bf 04} (2016) 017,
  [\href{http://arxiv.org/abs/1504.00659}{{\tt arXiv:1504.00659}}].

\bibitem{Junghans:2015hba}
D.~Junghans, {\it {Large-Field Inflation with Multiple Axions and the Weak
  Gravity Conjecture}},  \href{http://arxiv.org/abs/1504.03566}{{\tt
  arXiv:1504.03566}}.

\bibitem{Heidenreich:2015wga}
B.~Heidenreich, M.~Reece, and T.~Rudelius, {\it {Weak Gravity Strongly
  Constrains Large-Field Axion Inflation}},
  \href{http://arxiv.org/abs/1506.03447}{{\tt arXiv:1506.03447}}.

\bibitem{Palti:2015xra}
E.~Palti, {\it {On Natural Inflation and Moduli Stabilisation in String
  Theory}},  {\em JHEP} {\bf 10} (2015) 188,
  [\href{http://arxiv.org/abs/1508.00009}{{\tt arXiv:1508.00009}}].

\bibitem{Heidenreich:2015nta}
B.~Heidenreich, M.~Reece, and T.~Rudelius, {\it {Sharpening the Weak Gravity
  Conjecture with Dimensional Reduction}},  {\em JHEP} {\bf 02} (2016) 140,
  [\href{http://arxiv.org/abs/1509.06374}{{\tt arXiv:1509.06374}}].

\bibitem{Kooner:2015rza}
K.~Kooner, S.~Parameswaran, and I.~Zavala, {\it {Warping the Weak Gravity
  Conjecture}},  \href{http://arxiv.org/abs/1509.07049}{{\tt
  arXiv:1509.07049}}.

\bibitem{Kappl:2015esy}
R.~Kappl, H.~P. Nilles, and M.~W. Winkler, {\it {Modulated Natural Inflation}},
   \href{http://arxiv.org/abs/1511.05560}{{\tt arXiv:1511.05560}}.

\bibitem{Choi:2015aem}
K.~Choi and H.~Kim, {\it {Aligned natural inflation with modulations}},
  \href{http://arxiv.org/abs/1511.07201}{{\tt arXiv:1511.07201}}.

\bibitem{Ibanez:2015fcv}
L.~E. Ibanez, M.~Montero, A.~Uranga, and I.~Valenzuela, {\it {Relaxion
  Monodromy and the Weak Gravity Conjecture}},  {\em JHEP} {\bf 04} (2016) 020,
  [\href{http://arxiv.org/abs/1512.00025}{{\tt arXiv:1512.00025}}].

\bibitem{Hebecker:2015zss}
A.~Hebecker, F.~Rompineve, and A.~Westphal, {\it {Axion Monodromy and the Weak
  Gravity Conjecture}},  {\em JHEP} {\bf 04} (2016) 157,
  [\href{http://arxiv.org/abs/1512.03768}{{\tt arXiv:1512.03768}}].

\bibitem{Fonseca:2016eoo}
N.~Fonseca, L.~de~Lima, C.~S. Machado, and R.~D. Matheus, {\it {N-Relaxion}},
  \href{http://arxiv.org/abs/1601.07183}{{\tt arXiv:1601.07183}}.

\bibitem{Parameswaran:2016qqq}
S.~Parameswaran, G.~Tasinato, and I.~Zavala, {\it {Subleading Effects and the
  Field Range in Axion Inflation}},  {\em JCAP} {\bf 1604} (2016), no.~04 008,
  [\href{http://arxiv.org/abs/1602.02812}{{\tt arXiv:1602.02812}}].

\bibitem{Baume:2016psm}
F.~Baume and E.~Palti, {\it {Backreacted Axion Field Ranges in String Theory}},
   \href{http://arxiv.org/abs/1602.06517}{{\tt arXiv:1602.06517}}.

\bibitem{Garcia-Valdecasas:2016voz}
E.~García-Valdecasas and A.~Uranga, {\it {On the 3-form formulation of axion
  potentials from D-brane instantons}},
  \href{http://arxiv.org/abs/1605.08092}{{\tt arXiv:1605.08092}}.

\bibitem{Vafa:2005ui}
C.~Vafa, {\it {The String landscape and the swampland}},
  \href{http://arxiv.org/abs/hep-th/0509212}{{\tt hep-th/0509212}}.

\bibitem{Susskind:1995da}
L.~Susskind, {\it {Trouble for remnants}},
  \href{http://arxiv.org/abs/hep-th/9501106}{{\tt hep-th/9501106}}.

\bibitem{Banks:2010zn}
T.~Banks and N.~Seiberg, {\it {Symmetries and Strings in Field Theory and
  Gravity}},  {\em Phys. Rev.} {\bf D83} (2011) 084019,
  [\href{http://arxiv.org/abs/1011.5120}{{\tt arXiv:1011.5120}}].

\bibitem{Banks:2006mm}
T.~Banks, M.~Johnson, and A.~Shomer, {\it {A Note on Gauge Theories Coupled to
  Gravity}},  {\em JHEP} {\bf 09} (2006) 049,
  [\href{http://arxiv.org/abs/hep-th/0606277}{{\tt hep-th/0606277}}].

\bibitem{Nakayama:2015hga}
Y.~Nakayama and Y.~Nomura, {\it {Weak gravity conjecture in the AdS/CFT
  correspondence}},  {\em Phys. Rev.} {\bf D92} (2015), no.~12 126006,
  [\href{http://arxiv.org/abs/1509.01647}{{\tt arXiv:1509.01647}}].

\bibitem{Harlow:2015lma}
D.~Harlow, {\it {Wormholes, Emergent Gauge Fields, and the Weak Gravity
  Conjecture}},  {\em JHEP} {\bf 01} (2016) 122,
  [\href{http://arxiv.org/abs/1510.07911}{{\tt arXiv:1510.07911}}].

\bibitem{Horowitz:2016ezu}
G.~T. Horowitz, J.~E. Santos, and B.~Way, {\it {Evidence for an Electrifying
  Violation of Cosmic Censorship}},
  \href{http://arxiv.org/abs/1604.06465}{{\tt arXiv:1604.06465}}.

\bibitem{Benjamin:2016fhe}
N.~Benjamin, E.~Dyer, A.~L. Fitzpatrick, and S.~Kachru, {\it {Universal Bounds
  on Charged States in 2d CFT and 3d Gravity}},
  \href{http://arxiv.org/abs/1603.09745}{{\tt arXiv:1603.09745}}.

\bibitem{HarvardPaper}
B.~Heidenreich, M.~Reece, and T.~Rudelius, ``{Evidence for a Lattice Weak
  Gravity Conjecture}.'' To appear.

\bibitem{griffiths2009exact}
J.~Griffiths and J.~Podolsk{\`y}, {\em Exact Space-Times in Einstein's General
  Relativity}.
\newblock Cambridge Monographs on Mathematical Physics. Cambridge University
  Press, 2009.

\bibitem{Ortin:977337}
T.~Ortín, {\em {Gravity and strings}}.
\newblock Cambridge Univ. Press, Cambridge, 2004.

\bibitem{Penrose:1969pc}
R.~Penrose, {\it {Gravitational collapse: The role of general relativity}},
  {\em Riv. Nuovo Cim.} {\bf 1} (1969) 252--276. [Gen. Rel.
  Grav.34,1141(2002)].

\bibitem{Wald:1997wa}
R.~M. Wald, {\it {Gravitational collapse and cosmic censorship}},  in {\em {In
  *Iyer, B.R. (ed.) et al.: Black holes, gravitational radiation and the
  universe* 69-85}}, 1997.
\newblock \href{http://arxiv.org/abs/gr-qc/9710068}{{\tt gr-qc/9710068}}.

\bibitem{PhysRevD.14.870}
W.~G. Unruh, {\it Notes on black-hole evaporation},  {\em Phys. Rev. D} {\bf
  14} (Aug, 1976) 870--892.

\bibitem{Antoniadis:1997eg}
I.~Antoniadis, S.~Ferrara, R.~Minasian, and K.~S. Narain, {\it {R**4 couplings
  in M and type II theories on Calabi-Yau spaces}},  {\em Nucl. Phys.} {\bf
  B507} (1997) 571--588, [\href{http://arxiv.org/abs/hep-th/9707013}{{\tt
  hep-th/9707013}}].

\bibitem{Cheung:2014vva}
C.~Cheung and G.~N. Remmen, {\it {Naturalness and the Weak Gravity
  Conjecture}},  {\em Phys. Rev. Lett.} {\bf 113} (2014) 051601,
  [\href{http://arxiv.org/abs/1402.2287}{{\tt arXiv:1402.2287}}].

\bibitem{Deser:1983tn}
S.~Deser, R.~Jackiw, and G.~'t~Hooft, {\it {Three-Dimensional Einstein Gravity:
  Dynamics of Flat Space}},  {\em Annals Phys.} {\bf 152} (1984) 220.

\bibitem{Ibanez:2012zz}
L.~E. Ibanez and A.~M. Uranga, {\em {String theory and particle physics: An
  introduction to string phenomenology}}.
\newblock Cambridge University Press, 2012.

\bibitem{Denef:2008wq}
F.~Denef, {\it {Les Houches Lectures on Constructing String Vacua}},  in {\em
  {String theory and the real world: From particle physics to astrophysics.
  Proceedings, Summer School in Theoretical Physics, 87th Session, Les Houches,
  France, July 2-27, 2007}}, pp.~483--610, 2008.
\newblock \href{http://arxiv.org/abs/0803.1194}{{\tt arXiv:0803.1194}}.

\bibitem{Polyakov:1976fu}
A.~M. Polyakov, {\it {Quark Confinement and Topology of Gauge Groups}},  {\em
  Nucl. Phys.} {\bf B120} (1977) 429--458.

\bibitem{Affleck:1989qf}
I.~Affleck, J.~A. Harvey, L.~Palla, and G.~W. Semenoff, {\it {The
  {Chern-Simons} Term Versus the Monopole}},  {\em Nucl. Phys.} {\bf B328}
  (1989) 575--584.

\bibitem{Blumenhagen:2006xt}
R.~Blumenhagen, M.~Cvetic, and T.~Weigand, {\it {Spacetime instanton
  corrections in 4D string vacua: The Seesaw mechanism for D-Brane models}},
  {\em Nucl. Phys.} {\bf B771} (2007) 113--142,
  [\href{http://arxiv.org/abs/hep-th/0609191}{{\tt hep-th/0609191}}].

\bibitem{Ibanez:2006da}
L.~E. Ibanez and A.~M. Uranga, {\it {Neutrino Majorana Masses from String
  Theory Instanton Effects}},  {\em JHEP} {\bf 03} (2007) 052,
  [\href{http://arxiv.org/abs/hep-th/0609213}{{\tt hep-th/0609213}}].

\bibitem{Florea:2006si}
B.~Florea, S.~Kachru, J.~McGreevy, and N.~Saulina, {\it {Stringy Instantons and
  Quiver Gauge Theories}},  {\em JHEP} {\bf 05} (2007) 024,
  [\href{http://arxiv.org/abs/hep-th/0610003}{{\tt hep-th/0610003}}].

\bibitem{BerasaluceGonzalez:2011wy}
M.~Berasaluce-Gonzalez, L.~E. Ibanez, P.~Soler, and A.~M. Uranga, {\it
  {Discrete gauge symmetries in D-brane models}},  {\em JHEP} {\bf 12} (2011)
  113, [\href{http://arxiv.org/abs/1106.4169}{{\tt arXiv:1106.4169}}].

\bibitem{Banados:1992wn}
M.~Banados, C.~Teitelboim, and J.~Zanelli, {\it {The Black hole in
  three-dimensional space-time}},  {\em Phys. Rev. Lett.} {\bf 69} (1992)
  1849--1851, [\href{http://arxiv.org/abs/hep-th/9204099}{{\tt
  hep-th/9204099}}].

\bibitem{Carlip:1995qv}
S.~Carlip, {\it {The (2+1)-Dimensional black hole}},  {\em Class. Quant. Grav.}
  {\bf 12} (1995) 2853--2880, [\href{http://arxiv.org/abs/gr-qc/9506079}{{\tt
  gr-qc/9506079}}].

\bibitem{Birmingham:2001dt}
D.~Birmingham, I.~Sachs, and S.~Sen, {\it {Exact results for the BTZ black
  hole}},  {\em Int. J. Mod. Phys.} {\bf D10} (2001) 833--858,
  [\href{http://arxiv.org/abs/hep-th/0102155}{{\tt hep-th/0102155}}].

\bibitem{Carlip:2005zn}
S.~Carlip, {\it {Conformal field theory, (2+1)-dimensional gravity, and the BTZ
  black hole}},  {\em Class. Quant. Grav.} {\bf 22} (2005) R85--R124,
  [\href{http://arxiv.org/abs/gr-qc/0503022}{{\tt gr-qc/0503022}}].

\bibitem{Kraus:2006wn}
P.~Kraus, {\it {Lectures on black holes and the AdS(3) / CFT(2)
  correspondence}},  {\em Lect. Notes Phys.} {\bf 755} (2008) 193--247,
  [\href{http://arxiv.org/abs/hep-th/0609074}{{\tt hep-th/0609074}}].

\bibitem{Hyun:1994na}
S.~Hyun, G.~H. Lee, and J.~H. Yee, {\it {Hawking radiation from
  (2+1)-dimensional black hole}},  {\em Phys. Lett.} {\bf B322} (1994)
  182--187.

\bibitem{Ross:1992ba}
S.~F. Ross and R.~B. Mann, {\it {Gravitationally collapsing dust in
  (2+1)-dimensions}},  {\em Phys. Rev.} {\bf D47} (1993) 3319--3322,
  [\href{http://arxiv.org/abs/hep-th/9208036}{{\tt hep-th/9208036}}].

\bibitem{Andrade:2005ur}
T.~Andrade, M.~Banados, R.~Benguria, and A.~Gomberoff, {\it {The 2+1 charged
  black hole in topologically massive electrodynamics}},  {\em Phys. Rev.
  Lett.} {\bf 95} (2005) 021102,
  [\href{http://arxiv.org/abs/hep-th/0503095}{{\tt hep-th/0503095}}].

\bibitem{Coleman:1991ku}
S.~R. Coleman, J.~Preskill, and F.~Wilczek, {\it {Quantum hair on black
  holes}},  {\em Nucl. Phys.} {\bf B378} (1992) 175--246,
  [\href{http://arxiv.org/abs/hep-th/9201059}{{\tt hep-th/9201059}}].

\bibitem{Hawking:1982dh}
S.~W. Hawking and D.~N. Page, {\it {Thermodynamics of Black Holes in anti-De
  Sitter Space}},  {\em Commun. Math. Phys.} {\bf 87} (1983) 577.

\bibitem{Myung:2006sq}
Y.~S. Myung, {\it {Phase transition between the BTZ black hole and AdS space}},
   {\em Phys. Lett.} {\bf B638} (2006) 515--518,
  [\href{http://arxiv.org/abs/gr-qc/0603051}{{\tt gr-qc/0603051}}].

\bibitem{Eune:2013qs}
M.~Eune, W.~Kim, and S.-H. Yi, {\it {Hawking-Page phase transition in BTZ black
  hole revisited}},  {\em JHEP} {\bf 03} (2013) 020,
  [\href{http://arxiv.org/abs/1301.0395}{{\tt arXiv:1301.0395}}].

\bibitem{Hartman-lect}
T.~Hartman, {\it Lecture notes on quantum gravity and black holes},  2015.

\bibitem{Polchinskiv2}
J.~Polchinski, {\em String theory. Vol. 2: Superstring theory and beyond}.
\newblock Cambridge University Press, 1998.

\bibitem{Goddard:1988fw}
P.~Goddard and D.~I. Olive, {\it {KAC-MOODY AND VIRASORO ALGEBRAS: A REPRINT
  VOLUME FOR PHYSICISTS}},  {\em Adv. Ser. Math. Phys.} {\bf 3} (1988) 1--586.

\bibitem{Witten:1998qj}
E.~Witten, {\it {Anti-de Sitter space and holography}},  {\em Adv. Theor. Math.
  Phys.} {\bf 2} (1998) 253--291,
  [\href{http://arxiv.org/abs/hep-th/9802150}{{\tt hep-th/9802150}}].

\bibitem{Jensen:2010em}
K.~Jensen, {\it {Chiral anomalies and AdS/CMT in two dimensions}},  {\em JHEP}
  {\bf 01} (2011) 109, [\href{http://arxiv.org/abs/1012.4831}{{\tt
  arXiv:1012.4831}}].

\bibitem{Dijkgraaf:2000fq}
R.~Dijkgraaf, J.~M. Maldacena, G.~W. Moore, and E.~P. Verlinde, {\it {A Black
  hole Farey tail}},  \href{http://arxiv.org/abs/hep-th/0005003}{{\tt
  hep-th/0005003}}.

\bibitem{Keller:2014xba}
C.~A. Keller and A.~Maloney, {\it {Poincare Series, 3D Gravity and CFT
  Spectroscopy}},  {\em JHEP} {\bf 02} (2015) 080,
  [\href{http://arxiv.org/abs/1407.6008}{{\tt arXiv:1407.6008}}].

\bibitem{Schwimmer:1986mf}
A.~Schwimmer and N.~Seiberg, {\it {Comments on the N=2, N=3, N=4 Superconformal
  Algebras in Two-Dimensions}},  {\em Phys. Lett.} {\bf B184} (1987) 191--196.

\bibitem{Manschot:2008zb}
J.~Manschot, {\it {On the space of elliptic genera}},  {\em Commun. Num. Theor.
  Phys.} {\bf 2} (2008) 803--833, [\href{http://arxiv.org/abs/0805.4333}{{\tt
  arXiv:0805.4333}}].

\bibitem{Gaberdiel:2008xb}
M.~R. Gaberdiel, S.~Gukov, C.~A. Keller, G.~W. Moore, and H.~Ooguri, {\it
  {Extremal N=(2,2) 2D Conformal Field Theories and Constraints of
  Modularity}},  {\em Commun. Num. Theor. Phys.} {\bf 2} (2008) 743--801,
  [\href{http://arxiv.org/abs/0805.4216}{{\tt arXiv:0805.4216}}].

\bibitem{VanHerck:2009ww}
W.~Van~Herck and T.~Wyder, {\it {Black Hole Meiosis}},  {\em JHEP} {\bf 04}
  (2010) 047, [\href{http://arxiv.org/abs/0909.0508}{{\tt arXiv:0909.0508}}].

\bibitem{Hellerman:2010qd}
S.~Hellerman and C.~Schmidt-Colinet, {\it {Bounds for State Degeneracies in 2D
  Conformal Field Theory}},  {\em JHEP} {\bf 08} (2011) 127,
  [\href{http://arxiv.org/abs/1007.0756}{{\tt arXiv:1007.0756}}].

\bibitem{Keller:2013qqa}
C.~A. Keller, {\it {Modularity, Calabi-Yau geometry and 2d CFTs}},  {\em Proc.
  Symp. Pure Math.} {\bf 88} (2014) 307--316,
  [\href{http://arxiv.org/abs/1312.7313}{{\tt arXiv:1312.7313}}].

\bibitem{Hellerman:2009bu}
S.~Hellerman, {\it {A Universal Inequality for CFT and Quantum Gravity}},  {\em
  JHEP} {\bf 08} (2011) 130, [\href{http://arxiv.org/abs/0902.2790}{{\tt
  arXiv:0902.2790}}].

\bibitem{Beem:2014zpa}
C.~Beem, M.~Lemos, P.~Liendo, L.~Rastelli, and B.~C. van Rees, {\it {The $
  \mathcal{N}=2 $ superconformal bootstrap}},  {\em JHEP} {\bf 03} (2016) 183,
  [\href{http://arxiv.org/abs/1412.7541}{{\tt arXiv:1412.7541}}].

\bibitem{Cardy:1986ie}
J.~L. Cardy, {\it {Operator Content of Two-Dimensional Conformally Invariant
  Theories}},  {\em Nucl. Phys.} {\bf B270} (1986) 186--204.

\end{thebibliography}\endgroup

\end{document}